\documentclass[default,iicol]{sn-jnl}

\usepackage{multirow}
\usepackage{amssymb,amsmath,amsthm,xcolor}
\usepackage{amsfonts}
\usepackage{graphicx,psfrag,epsf}
\usepackage{enumerate}
\usepackage{hyperref}
\usepackage[normalem]{ulem}
\usepackage{caption}
\usepackage{subcaption}
\usepackage{multicol}
\usepackage{natbib}
\setcitestyle{authoryear,open={(},close={)}} 

\jyear{2022}

\theoremstyle{thmstyleone}%
\newtheorem{theorem}{Theorem}
\newtheorem{corollary}{Corollary}
\newtheorem{lemma}{Lemma}%

\theoremstyle{thmstylethree}%

\DeclareMathOperator{\Var}{Var}

\newcommand{\comment}[1]{}

\begin{document}

\title[SURE-tuned Bridge Regression]{SURE-tuned Bridge Regression}


\author*[1]{\fnm{Jorge} \sur{Lor\'ia}}\email{loria@purdue.edu}

\author[2]{\fnm{Anindya} \sur{Bhadra}}

\affil[1]{\orgdiv{Department of Statistics}, \orgname{Purdue University}, \orgaddress{ \city{West Lafayette}, \postcode{47907}, \state{Indiana}, \country{United States}}}


\abstract{Consider the \unboldmath{$\ell_{\alpha}$} regularized linear regression, also termed Bridge regression. For $\alpha\in (0,1)$, Bridge regression enjoys several statistical properties of interest such as sparsity and near-unbiasedness of the estimates \citep{FanLiSCADJASA}. However, the main difficulty lies in the non-convex nature of the penalty for these values of $\alpha$, which makes an optimization procedure challenging and usually it is only possible to find a local optimum. To address this issue, \citet{PolsonScottBayesBridge} took a sampling based fully Bayesian approach to this problem, using the correspondence between the Bridge penalty and a power exponential prior on the regression coefficients. However, their sampling procedure relies on Markov chain Monte Carlo (MCMC) techniques, which are inherently sequential and not scalable to large problem dimensions. Cross validation approaches are similarly computation-intensive. To this end, our contribution is a novel \emph{non-iterative} method to fit a Bridge regression model. The main contribution lies in an explicit formula for Stein's unbiased risk estimate for the out of sample prediction risk of Bridge regression, which can then be optimized to select the desired tuning parameters, allowing us to completely bypass MCMC as well as computation-intensive cross validation approaches. Our procedure yields results in a fraction of computational times compared to iterative schemes, without any appreciable loss in statistical performance. An \texttt{R} implementation is publicly available online at: \href{ https://github.com/loriaJ/Sure-tuned_BridgeRegression}{https://github.com/loriaJ/Sure-tuned_BridgeRegression}.
}




\maketitle

\section{Introduction}\label{sec1}

For regression coefficients $\beta\in \mathbb{R}^p$, the Bridge regression estimate is usually stated as the solution to the following optimization problem:
\begin{align}
    \hat\beta = \arg\min_{\beta}\left\{\frac{1}{2}\lVert y - X\beta\rVert_2^2 + \nu\sum_{i}\lvert\beta_i\rvert^\alpha\right\},\label{eq:bridge}
\end{align}
where $y \in \mathbb{R}^n$ is the response variable, $X \in \mathbb{R}^{n\times p}$ is the design matrix, $\nu>0$ is a penalty parameter, and $\alpha\in(0,2]$ is the coefficient exponent. Particular cases of interest are: the degenerate $\alpha=0$, which corresponds to the $\ell_0$ penalized regression; when $\alpha=1$, it corresponds to the lasso procedure; and $\alpha=2$ is ridge regression. 
When $\alpha\in(0,1)$, this optimization problem is non-convex, and for $\alpha \in[1,2]$, it is a convex problem. Furthermore, the Bridge model has the desirable properties of sparsity and near-unbiasedness \citep{FanLiSCADJASA}, when $\alpha \in (0,1).$

However, for the same setting of $\alpha\in (0,1)$, \citet{PolsonScottBayesBridge}
argue that a purely optimization-based approach is inappropriate, as the penalized likelihood surface is multi-modal, which in turn leads to several possible solutions or local optimality conditions instead of global optimality, which is hard to establish \citep{mazumder2011sparsenet}. In this context, three iterative strategies are currently available for solving this problem, to the best of our knowledge. These are: 

\begin{enumerate}
    \item  A sampling based fully Bayesian Markov chain Monte Carlo (MCMC) procedure by  \citet{PolsonScottBayesBridge}. The key to their approach is the observation that a solution to the optimization problem in Equation~\eqref{eq:bridge} could be obtained as the posterior mode under the model:
    \begin{eqnarray*}
    y\mid X,\beta &\sim& \mathcal{N}(X\beta,\sigma^2),\\
    p(\beta_i) &\propto& \exp(-\nu \lvert \beta_i \lvert^\alpha),
    \end{eqnarray*}
    where the density on the last line was termed the exponential power density by \citet[][p.~157]{BoxTiao1973}. Consequently, \citet{PolsonScottBayesBridge} recommend a fully Bayesian approach using this hierarchical model and outline  two possible sampling schemes. Also of interest are the approaches of  \citet{GomezSanchezBridge}, and \citet{MallickYiBridge}.
    
    \item Expectation-maximization (EM) point estimation routines for finding the maximum a posteriori estimate under a power exponential prior, which coincides to the solution of the optimization problem in Equation~\eqref{eq:bridge}, proposed by  \citet{PolsonScottEM}, and \citet{MallickYiBridge}. Moreover, pathwise coordinate descent approaches are also available for solving the penalized optimization  problem \citep{mazumder2011sparsenet,griffin2022improved, MarjanovicSolo}.
    
    \item Finally, a variational Bayes approach, for a tractable approximation to the target posterior under a fully Bayesian model  \citep{ArmaganBridge}.
\end{enumerate}
While the approaches above could be termed respectively as fully Bayesian, frequentist and approximately Bayesian; and hence together they cover the full spectrum of statistical inference reasonably well, a common and recurring theme is their reliance on iterative routines for the purpose of model fitting, inherently limiting their scalability. The main contribution of the current paper is to present a \emph{non-iterative} approach. Our approach exploits a closed form expression for the desired posterior moments in a penalized likelihood formulation of the Bridge problem, given a latent Gaussian representation, using the celebrated result of \citet{West1987} that an exponential power density, $p_X(x)\propto \exp(-\lvert x\lvert^\alpha)$, for $\alpha\in (0,1)$, is a normal scale mixture with respect to a positive $\alpha/2$ stable density for the latent scale variable. This result leads to a closed form expression for Stein's unbiased risk estimate or SURE \citep{stein_inadmissibility_1956, stein1981estimation} for Bridge regression, allowing a selection of $\nu$ by minimizing SURE via a simple one-dimensional grid search. Upon selecting $\nu$, the penalized likelihood estimate of $\beta$ that coincides with the Bridge optimization problem is also available analytically, once again by exploiting the latent Gaussian representation. The connection between SURE and cross validation has been explored by \citet{Efron2004}, who showed the latter to be a Monte Carlo estimate of SURE. Thus, choosing the desired tuning parameters via minimizing SURE allows us to completely bypass computationally demanding cross validation procedures. Another notable method to avoid the high computational burden of cross validation is an approximate leave-one-out cross validation due to \citet{WangALOCV}. However, its theoretical investigations are currently restricted to convex penalties. The explicit formula resulting from our latent Gaussian representation preempts iterative Markov chain Monte Carlo simulations as well, since the desired posterior moments are available in closed form. Our results still require numerical evaluation by means of vanilla Monte Carlo of some functions of the latent stable scale variable, but these evaluations are not inherently sequential in the same sense an MCMC or an iterative optimization routine is.

We compare our procedure to the fully Bayesian approach of \citet{PolsonScottBayesBridge}, which is implemented in the \texttt{R} package \texttt{BayesBridge} as well as with cross validation using an EM approach, also implemented in the \texttt{R} package \texttt{BayesBridge}. Our results indicate similar statistical performance, but with a run time that is typically much faster than a fully Bayesian MCMC routine or cross validation. Predictive comparison on a set of spectral reflectances for photosynthetic prediction in plants \citep{MeachamHensoldMontesWuetal} is also presented, once again yielding similar out-of-sample prediction errors, but in a fraction of the time.

To summarize, our main contributions are: 
\begin{enumerate}
    \item An explicit formula for the desired posterior moments in a latent Gaussian representation of the Bridge regression model, yielding a closed form expression for SURE. 
    \item A demonstration that  the estimates have finite Monte Carlo variance, while avoiding iterative MCMC and optimization routines altogether.
    \item A numerical demonstration on simulated and a  photosynthetic data set, indicating similar statistical performance with a large saving in computational time.
\end{enumerate}

The rest of the manuscript is organized as follows. In Section \ref{sec:method} we present explicit formulas for the Bridge regression model. Next, in Section~\ref{sec:SURE} we include a closed form expression for $\mathrm{SURE}$, for a general prior on the coefficients of a linear regression. We also provide details of the implementation that makes clear how we are able to avoid an iterative routine, with vanilla Monte Carlo being sufficient for our purposes. This is followed by Section~\ref{sec:Results} where we consider simulation experiments to validate our method, and verify the bounds on the variances. Further, in Section~\ref{sec:real_data} we implement our method in a prediction problem with spectroscopic measurements. We conclude in Section~\ref{sec:Conclusion} by pointing out some future directions.

\section{A Penalized Likelihood Framework for Bridge Regression}\label{sec:method}
In what follows, consider $\alpha\in(0,2)$, a fixed constant. Denote a random variable $B$ following the exponential power distribution as $B\sim EP(\alpha,\nu)$, with probability density function:
\begin{align*}
    p_\nu(\beta) = &\frac{\alpha\nu
    }{\Gamma(1/\alpha)}\exp\left(-2^\alpha\nu^{\alpha}\lvert\beta\rvert^\alpha\right).
\end{align*}
There is a direct relationship between the density of the exponential power distribution, and the penalty we are considering, since $-\log(p_\nu(\beta)) =  2^\alpha\nu^\alpha\lvert\beta\rvert^\alpha$, up to an additive constant. That is, the penalty is the negative logarithm of the density function, for a fixed parameter $\nu$. With that in mind, and motivated by the observations of \citet{West1987} and \citet{PSLocalShrinkage}, we consider a positive $\alpha/2-$stable random variable $L\sim S^+(\alpha/2,\nu)$, usually defined by its Laplace transform, given by: $\psi_L(\zeta)=\mathbb{E}[\exp(-\zeta L)]=\exp(-\nu^{\alpha/2} \lvert \zeta\rvert^{\alpha/2})$, \citep[see Equation 4.5,][]{ContTankov}. This is because explicit analytical expressions for the density of a positive $\alpha$-stable variable exist only for certain special cases. Making use of the Laplace exponent, \citet{West1987} and \citet{PSLocalShrinkage} expressed the density of an exponential power as a normal mixture of positive stable densities. Namely:
\begin{align}
    p_\nu(\beta) \propto&  \psi_L(\beta)
    =\int_0^\infty\exp\Big(-\frac{\beta^2x}{2\nu}\Big)p_L(x)dx, \label{eq:StableNormalMixture}
\end{align}
where we denote by $p_L(x)$ the density of  a positive $\alpha/2-$stable random variable with scale parameter $1$, denoted as $S^+(\alpha/2,1)$.
Noting that the exponential term is proportional to a mean zero normal density in $\beta$  with variance $\nu x^{-1}$, a multiplicative factor of $\nu^{-1/2} x^{1/2}$ is needed to complete the normal density. This motivates using the polynomially-tilted positive stable density: $T \sim PS^+(\alpha/2,\delta),\; \delta\ge 0$, with density $p_T$, given by $p_T(x)\propto x^{-\delta}p_L(x)$, where $p_L$ is the density of $L\sim S^+(\alpha/2,1)$, defined by \citet{devroye2009polynomiallytiltedStable}. We obtain:
\begin{align*}
p_\nu(\beta)\propto & \int_{0}^{\infty} p_\nu(\beta\mid x)x^{-1/2}p_L(x)dx\\
\propto & \int_{0}^{\infty} p_\nu(\beta\mid x)p_T(x)dx,
\end{align*}
where $p_\nu(\beta\mid x)= \mathcal{N} (\beta\mid 0, \nu x^{-1})$ and $T\sim PS^+(\alpha/2,1/2)$. The previous expression shows that the density of an exponential power acts as the marginal of a normal mixture model. To make use of this for the Bridge model, we apply it to the prior and obtain the posterior moments of $\beta$ conditional on $T$ and $y$. The following hierarchy corresponds to the $n$-means model:
\begin{align}
y_i\mid\beta_i \stackrel{ind}\sim & \mathcal{N}(\beta_i,\sigma^2), \label{eq:likelihood_n_means}\\
\beta_i \overset{iid}{\sim}& EP(\alpha,\nu),\label{eq:prior_density_n_means}
\end{align}
\sloppy for $i=1,\dots,n$, and 
$\sigma^2\in\mathbb{R}^+$. To obtain the posterior estimates in this $n$-means model, we make use of the following lemma.
\begin{lemma}
\label{le:equivalence_between_models_n_means}
The Bridge $n$-means model, given by Equations \eqref{eq:likelihood_n_means} and \eqref{eq:prior_density_n_means}, is equivalent to the hierarchical model:
\begin{align*}
    y_i\mid \beta_i,T_i\stackrel{ind}\sim\,& \mathcal{N}(\beta_i,\sigma^2),\\
    \beta_i\mid T_i\stackrel{ind}\sim& \mathcal{N}(0,\nu T_i^{-1}),\\
    T_i  \overset{i.i.d.}{\sim}\,& PS^+(\alpha/2,1/2),
\end{align*}
for $i=1,\dots,n$.
\end{lemma}
Proof of Lemma~\ref{le:equivalence_between_models_n_means} can be found in Appendix~\ref{pf:equivalence_between_models_n_means}. On its own, the previous lemma is not very surprising and is an immediate consequence of the result by \citet{West1987}. Its usefulness becomes clear in the following theorem.
\begin{theorem}\label{th:n_means_moments}
Under the model defined by Equations~\eqref{eq:likelihood_n_means} and~\eqref{eq:prior_density_n_means}, the marginal density for $y_i$ is:
\begin{align*}
    m_\nu(y_i) =& \int_{\mathbb{R}^+}p_\nu(y_i\mid t)p_T(t)dt < \infty, 
\end{align*}
and the posterior expectation of the first two moments of $\beta_i$ is given by: 
\begin{align*}
\tilde{\beta_i} :=&
    \mathbb{E}[\beta_i \mid y_i]\\
     =&  \frac{1}{m_\nu(y_i)}\int_{0}^\infty  y_i\nu(\sigma^2t+\nu)^{-1} p_\nu(y_i\mid t)p_T(t)dt ,\\
    \tilde{\beta}_i^{(2)}
    = & \mathbb{E}[\beta_i^2 \mid y_i]\\
    = & \frac{1}{m_\nu(y_i)}\int_{0}^\infty \left(\frac{\sigma^2\nu}{\sigma^2t+\nu} +\frac{y_i^2\nu^2}{(\sigma^2t+\nu)^{2}}\right)\\ &\hspace{1.3cm} \times p_\nu(y_i\mid t)p_T(t)dt, 
    \end{align*}
    where the densities in the integrals correspond to ${T\sim PS^+(\alpha/2,1/2)}$, and $y_i\mid t\sim \mathcal{N}(0,\sigma^2 + \nu t^{-1})$ .
\end{theorem}
Proof of Theorem~\ref{th:n_means_moments} can be found in Appendix~\ref{pf:n_means_moments}. Theorem~\ref{th:n_means_moments} implies that we can estimate the marginal density, $m_\nu(y_i)$, through a Monte Carlo (and not MCMC) averaging, by sampling from the distribution of $T$. This sampling is easily done by following the method proposed in \citet{devroye2009polynomiallytiltedStable}. We are not specifically interested in estimating the marginal density. More importantly, we can estimate the posterior expectation and variance of $\beta_i$ through a Monte Carlo simulation.  
Specifically, let $T_1,\dots,T_J\overset{iid}{\sim}PS^+(\alpha/2,1/2)$, and denoting $\mathcal{T}=(T_1,\dots,T_J)$, define the estimates as: 
\begin{align*}
    m_\nu(y_i)_\mathcal{T} & =\frac{1}{J}\sum_{j=1}^J p_\nu(y_i\mid T_j), \\
    \mathbb{E}[\beta_i\mid y_i]_\mathcal{T} & = \frac{1}{J m_\nu(y_i)_\mathcal{T}}\sum_{j=1}^Jp_\nu(y_i\mid T_j)\mathbb{E}[\beta_i\mid y,T_j],\\
    \mathbb{E}[\beta_i^2\mid y_i]_{\mathcal{T}} & = \frac{1}{J m_\nu(y_i)_\mathcal{T}}\sum_{j=1}^Jp_\nu(y_i\mid T_j)\left(\Var[\beta_i\mid y_i,T_j]\right. \\
    & \left.\hspace{3cm}+ \mathbb{E}[\beta_i\mid y_i,T_j]^2\right).
\end{align*}
In a Monte Carlo estimation, it is natural to wonder: (1) whether the estimates are unbiased and (2) whether they have bounded variance. Now we present a small technical lemma which gives a bound for the marginal. This helps us address the above concerns.
\begin{lemma}\label{le:marginal_bounded}
The marginal under the $n$-means model admits the following lower bound:
\begin{align*}
    m_\nu(y_i) > \exp\left(-\frac{y_i^2}{2\sigma^2}\right)C_{(\sigma^2,\nu)},
\end{align*}
where $C_{(\sigma^2,\nu)}$ is a strictly positive constant independent of $y$.
\end{lemma}
Proof of Lemma~\ref{le:marginal_bounded} can be found in Appendix~\ref{pf:bounded_marginal}. The two concerns above arise since (1)~we are using a ratio of two Monte Carlo estimates that come from the same simulations and (2) we employ simulations of random variables which are related to
$\alpha-$stable densities. The following theorem settles those concerns.
\begin{theorem}\label{th:finite_var}
Our estimators have the following properties:
\begin{enumerate}
    \item The estimator $m_\nu(y_i)_\mathcal{T}$ un-biasedly estimates $m_\nu(y_i)$. Next, $\mathbb{E}[\beta_i\mid y_i]_\mathcal{T}$ is an asymptotically unbiased estimator of $\tilde\beta_i$, as well as $\mathbb{E}[\beta_i^2\mid y]_\mathcal{T}$ of $\tilde\beta_i^{(2)}$, when $J\to\infty$. 

\item The variances of our Monte Carlo estimates are finite and are bounded by:
\begin{align*}
    \Var[m_\nu(y_i)_\mathcal{T}\mid y_i] &< J^{-1}(2\pi\sigma^2)^{-1}, \\
    \Var[\mathbb{E}(\beta_i\mid y)_{\mathcal{T}}\mid y_i] &< J^{-1} (2\pi\sigma^2)^{-1}\exp(y_i^2/\sigma^2)\times\\
    & C_{(\sigma^2,\nu)}^{-2}(\lvert y_i\rvert+ \tilde{\beta_i})^2,\text{ and}\\
    \Var[\mathbb{E}(\beta_i^2\mid y_i)_{\mathcal{T}}\mid y_i] &<
    J^{-1}(2\pi\sigma^2)^{-1}\exp(y_i^2/\sigma^2)\times \\
    & C_{(\sigma^2,\nu)}^{-2}(\sigma^2 + y_i^2-\tilde{\beta}^{(2)})^2. 
\end{align*}
\end{enumerate}
\end{theorem}
Proof of Theorem~\ref{th:finite_var} can be found in Appendix~\ref{pf:th_finite_var}. The first part of the theorem ensures our estimates are asymptotically unbiased. The second part of this theorem is also non-trivial, for two reasons: (1) the inverse of the marginal has been shown to have infinite variance in some circumstances \citep{newton1994approximate}, and (2) positive $\alpha/2-$stable do not even have a finite first moment \citep[Property 1.2.16]{SamorodnitskyTaqqu}. This means some care must be exercised when working with variables related to them. Having a finite variance ensures convergence to a normal distribution for our estimator at the usual parametric rate $J^{-1/2}$ by the central limit theorem. It also bears mention that our approximation of the desired posterior moments for $\beta_i$ consists of approximating the numerator and the denominator (the marginal $m(y_i)$) separately via vanilla Monte Carlo by drawing from the prior of $T$ that can be naively vectorized. It is certainly possible to recast the desired moments with respect to $(T\mid y)$. However, i.i.d. draws from this posterior are not available directly, and would necessitate an iterative MCMC technique, something we strain to avoid in the current work.

Now, we continue to explore the general case, where we have a vector of $n$ observations $y$, with design matrix $X$ with $p$ covariates. We do not impose any conditions on $n$ or $p$, other than these being positive integers. That is, we contemplate both $n\geq p$ and $p>n$ cases. The likelihood and priors are given by Equations  \eqref{eq:multivariate_likelihood} and \eqref{eq:multivariate_prior}. Namely:
\begin{align}
    y\mid X,\beta\sim& \mathcal{N}(X\beta,\Sigma),\label{eq:multivariate_likelihood}\\
    \beta_i\overset{iid}{\sim}&EP(\alpha,\nu),\text{ for } i = 1,\dots,p.\label{eq:multivariate_prior}
\end{align}
Similar to Lemma~\ref{le:equivalence_between_models_n_means}, we present Lemma~\ref{le:equivalence_between_models}, where we use the mixture representation of the exponential power to derive an equivalence between the Bridge model and a normal hierarchical model. 
This equivalence is in the sense that the $T_i$ act as lurking variables, and by integrating them we recover the marginal prior, the negative of the logarithm of which is the desired Bridge penalty. 
\begin{lemma}\label{le:equivalence_between_models}
The Bridge regression model, given by Equations \eqref{eq:multivariate_likelihood} and \eqref{eq:multivariate_prior}, is equivalent to the hierarchical model:
\begin{align*}
    y\mid X,\beta,T\sim\,& \mathcal{N}(X\beta,\Sigma),\\
    \beta_i\mid T_i\overset{ind}{\sim}& \mathcal{N}(0,\nu T_i^{-1}),\\ 
    T_i  \overset{iid}{\sim}\,& PS^+(\alpha/2,1/2), 
\end{align*}
for $i=1,\dots,p$.
\end{lemma}
The proof of Lemma~\ref{le:equivalence_between_models} can be found in Appendix~\ref{pf:equivalence_between_models}.
Lemma~\ref{le:equivalence_between_models} links the Bridge model to a normal hierarchical model, where the prior on the variance of the coefficients is given by a polynomially-tilted stable distribution. In what follows, we prove results for the Bridge model, going through the hierarchical model stated in Lemma~\ref{le:equivalence_between_models}. These results give a simple way to compute the marginal of $y$ and the posterior mean and variance of the coefficients. Although we make use of the marginal, we only need its value up to a normalizing constant. This avoids the complications that arise from estimation of marginal likelihood.

\begin{theorem}\label{th:multivariate_case}
Under the model given by Equations~\eqref{eq:multivariate_likelihood} and \eqref{eq:multivariate_prior}, the marginal density of $y$ is given by:
\begin{align*}
     m_\nu(y) = & \int_{(\mathbb{R}^{+})^p} p_\nu(y\mid \mathbf{t})\prod_{i=1}^p p_T(t_i)dt_i < \infty, 
\end{align*}
where $\mathbf{t}=(t_1,\dots,t_p)$.
We can compute the posterior first and second moments of $\beta$, and these are respectively given by:
\begin{align*}
    \tilde{\beta}=&\mathbb{E}_\nu[\beta\mid y]\\
    = & \frac{1}{m_\nu(y)}\int_{(\mathbb{R}^{+})^p}p_\nu(y\mid \mathbf{t}) \mathbb{E}_\nu[\beta\mid y,\mathbf{t}]\\ &\hspace{1cm}\times\prod_{i=1}^p p_T(t_i)dt_i,\\
    \tilde{\beta}^{(2)}=&\mathbb{E}_\nu(\beta \beta^T\mid y)\\ 
    = &\frac{1}{m_\nu(y)}\int_{(\mathbb{R}^{+})^p}p_\nu(y\mid \mathbf{t}) \mathbb{E}_\nu(\beta\beta^T\mid y,\mathbf{t})\\ & \hspace{1cm}\times \prod_{i=1}^p p_T(t_i)dt_i,
\end{align*}
where $p_\nu(y\mid \mathbf{t})$ is the density: $y\mid \mathbf{t}\sim\mathcal{N}_n(0,\nu V_{\mathbf{t},\nu})$, and the expectations inside the integrals come from: $\beta\mid y,\mathbf{t}\sim \mathcal{N}_p(\Lambda_\mathbf{t}X^TV_{\mathbf{t},\nu}^{-1}y,\nu\Lambda_\mathbf{t} - \nu\Lambda_\mathbf{t}X^TV_{\mathbf{t},\nu}^{-1}X\Lambda_\mathbf{t})$. Denoting $\Lambda_\mathbf{t} = \mathrm{diag}(t_i^{-1}); i=1,\dots,p$, $V_{\mathbf{t},\nu}=(X\Lambda_\mathbf{t} X^T + \nu^{-1}\Sigma)$.
\end{theorem}

The proof can be found in Appendix~\ref{pf:multivariate_lemma}. 
We omit the normalizing constants in Theorem~\ref{th:multivariate_case} as those appear in the numerator and denominator. As previously mentioned, this result holds for both $n\geq p$ and $n< p$, since we do not need the inverse of $X^TX$. It is possible to include a covariance structure on the observations if one is known beforehand. Also, the constant $\nu$ acts as a  balance between the sample covariance matrix of the covariates: $\nu X\Lambda_\mathbf{t} X^T$ and the error variance given by $\Sigma$, with $\nu$ implicitly weighing between these two variances. Furthermore, in Corollary~\ref{cor:simpler_formulas} we give an explicit expression for the first two moments of the fitted values. 
\begin{corollary}\label{cor:simpler_formulas}
The posterior expectation and variance of $X\beta$ given $y,\mathbf{t}$ can be expressed as:
\begin{align*}
    \mathbb{E}[X\beta \mid y,\mathbf{t} ]= A_\mathbf{t}(A_\mathbf{t}+\nu^{-1}\Sigma)^{-1}y,\\
    \Var(X\beta\mid y,\mathbf{t}) = \nu\Sigma(A_\mathbf{t}+\nu^{-1}\Sigma)^{-1}A_\mathbf{t},
\end{align*}
where we define $A_\mathbf{t} = X\Lambda_\mathbf{t}X^T$.
\end{corollary}
Proof of Corollary~\ref{cor:simpler_formulas} can be found in Appendix~\ref{pf:simpler_formulas}. This corollary is relevant since we can compute $A_\mathbf{t}$ after simulating $\Lambda_\mathbf{t}$. Once that is done, we just have to perform inversion and addition to compute the posterior mean and variance of the observations, for a fixed $\nu$. This is once again plain Monte Carlo that can be vectorized and not MCMC, allowing us to bypass an iterative routine resulting in faster computation.

Using now $\mathcal{T}=\{\mathbf{T}_{j}: j=1,\dots J\}$, where for each $j$, $\mathbf{T}_{j}=(T_{1,j},\dots,T_{p,j})$, and  ${T_{i,j}\overset{iid}{\sim}PS^+(\alpha/2,1/2)}$, we define the Monte Carlo estimates:
\begin{align*}
    m_\nu(y)_\mathcal{T} & =\frac{1}{J}\sum_{j=1}^J p_\nu(y\mid \mathbf{T}_j), \\
    \mathbb{E}[\beta\mid y]_\mathcal{T} & = \frac{1}{J m_\nu(y)_\mathcal{T}}\sum_{j=1}^Jp_\nu(y\mid \mathbf{T}_j)\mathbb{E}[\beta\mid y,\mathbf{T}_j],\\
    \mathbb{E}[\beta\beta^T\mid y]_{\mathcal{T}} & = \frac{1}{Jm_\nu(y)_\mathcal{T}}\sum_{j=1}^Jp_\nu(y\mid \mathbf{T}_j)\left\{\Var[\beta\mid y,\mathbf{T}_j]\right. \\
    & \left.\hspace{2cm}+ \mathbb{E}[\beta\mid y,\mathbf{T}_j]\mathbb{E}[\beta\mid y,\mathbf{T}_j]^T\right\}.
\end{align*}
As before, we want to address the unbiasedness and finite variance of these estimators. However, first we need a small technical lemma.
\begin{lemma}\label{le:bounded_marginal_vector}
The marginal  $m_\nu(y)$ in the linear regression setting admits the following lower bound:
\begin{align*}
    m_\nu(y) > \exp(-y^T\Sigma^{-1}y/2)C_{(\Sigma,\nu,X)},
\end{align*}
where $C_{(\Sigma,\nu,X)}$ is a strictly positive constant independent of $y$. 
\end{lemma}
Proof of Lemma~\ref{le:bounded_marginal_vector} can be found in Appendix~\ref{pf:bounded_marginal_vector}. This lemma is a building block for the bounds of the variances that we give in the following theorem.
\begin{theorem}\label{th:finite_var_vector}
Our estimators have the following properties:

\begin{enumerate}
    \item The estimator $m_\nu(y)_\mathcal{T}$ unbiasedly estimates the marginal $m_\nu(y)$. Next, $\mathbb{E}[\beta\mid y]_\mathcal{T}$ and ${\mathbb{E}[\beta\beta^T\mid y]_\mathcal{T}}$ are asymptotically unbiased estimators of $\tilde{\beta}$, and $\tilde\beta^{(2)}$. 
    
    \item The variances of these Monte Carlo estimates are finite and have the following explicit bounds:
\begin{align*}
    \mathrm{Var}[m_\nu(y)_\mathcal{T}\mid y] & < J^{-1}(2\pi)^{-p}\det(\Sigma^{-1}),\\
    \Var(\lVert \mathbb{E}[\beta\mid y]_{\mathcal{T}}\rVert_2\mid y) &< J^{-1}(2\pi)^{-p}\det(\Sigma^{-1})\\
    & \times \exp(y^T\Sigma^{-1}y)C_{(\Sigma,\nu,X)}^{-2}\\
    &\times (\nu^2 pK_2 M\\
    & + \lVert\tilde\beta\rVert^2_2 +2p \nu K_1\sqrt{M}\lVert\tilde{\beta}\rVert_2),\\
    \Var\left(\lVert \mathbb{E}[\beta\beta^T\mid y]_{\mathcal{T}}\rVert_2 \mid y\right) & < J^{-1}(2\pi)^{-p}\det(\Sigma^{-1})\\
    & \times \exp(y^T\Sigma^{-1}y)C_{(\Sigma,\nu,X)}^{-2}\\
    & \times (C +2M_2\lVert\tilde{\beta}^{(2)}\rVert_2 +\\
    &+\lVert\tilde{\beta}^{(2)}\rVert^2_2)
\end{align*}

where $C=p K_2 + 2Mp K_3 + M\frac{p(p-1)}{2}K_2K_1
+M^2K_4+M^2p(p-1)K_2^2$, $M_2 = \nu p K_1+MpK_2$, $K_i=\int_{\mathbb{R}^+}t^{-i}p_T(t)dt$, and $M=\lVert\Sigma^{-1}\rVert^2_2\lVert X\rVert^2_2 \lVert y\rVert^2_2$. 
\end{enumerate}
\end{theorem}
The proof of the previous theorem can be found in Appendix~\ref{pf:finite_var_vector}.

Although Theorems~\ref{th:multivariate_case} and~\ref{th:finite_var_vector} extend the results for the $n$-means models to a regression setting, it still is of interest to understand how the $n$-means case works in practice, as has been done before for other non-convex regression approaches, for example, the horseshoe regression \citep{BhadraJMLR}. Denote $r=\min(n,p)$, and consider the singular value decomposition of $X=UDV^T$, where $U,D,V$ are matrices with dimensions: $n\times r$,\;$r\times r$, and $p\times r$, respectively, $D$ is a diagonal matrix with positive entries, $U$ and $V$ satisfy: $U^TU=V^TV=I_r$, with $I_r$ the $r\times r$ identity matrix. Based on these definitions, we further define: $Z=UD$, and place the prior on a linear transformation of $\beta$ given by $V\gamma$, instead of on $\beta$. Explicitly, the model we use is:
\begin{align}
    y\mid X,\beta &\sim \mathcal{N}(X\beta,\sigma^2I_n),\label{eq:multivariate_normal}\\
    \beta &= V\gamma,\label{eq:beta_gamma_relationship}\\
    \gamma_i &\overset{iid}{\sim} EP(\alpha,\nu)\text{, for } i = 1,\dots,p.\label{eq:gamma_priors}
\end{align}
As an immediate consequence, the least squares estimate of $\gamma$ is given by: $\hat \gamma = (Z^TZ)^{-1}Z^Ty$. Then, $\hat \gamma\mid \gamma \sim \mathcal{N}(\gamma,\sigma^2D^{-2})$, which has diagonal variance matrix with positive entries, by definition of $D$. That is, we have reduced this case to Theorem~\ref{th:n_means_moments}. We formalize this in Corollary~\ref{cor:svd_decomp}.

\begin{corollary}\label{cor:svd_decomp}
    Under Equations~\eqref{eq:multivariate_normal}, \eqref{eq:beta_gamma_relationship} and \eqref{eq:gamma_priors}, 
    \begin{align*}
        \mathbb{E}[\beta\mid \hat\gamma,\nu,\sigma^2]=V^T\mathbb{E}[\gamma\mid \hat\gamma,\nu,\sigma^2D^{-2}],
    \end{align*} 
    where the marginal of $\hat\gamma_i$ is given in Theorem \ref{th:n_means_moments}, with variance $\sigma^2d_i^{-2}$. The posterior variance and posterior expectation of $\gamma$ can be computed using Theorem \ref{th:n_means_moments}.
\end{corollary}
The proof of Corollary \ref{cor:svd_decomp} can be found in Appendix~\ref{pf:svd_decomp}. We can now estimate the posterior mean and variance for a fixed penalty $\nu$, in both the general model and the $n$-means model, using the SVD. 

\section{SURE for Bridge Regression}\label{sec:SURE}
A central question in Bridge regression, or for that matter, in any penalized regression, is how to choose the penalty parameter $\nu$ in Equation~\eqref{eq:bridge}. While the closed form expression and Monte Carlo estimates of the posterior moments of $\beta$ have been worked out in the previous section, the motivation has not been quite clear yet. In this section, we demonstrate that we now have all the necessary ingredients for computing Stein's unbiased risk estimate or SURE \citep{stein_inadmissibility_1956,stein1981estimation} for Bridge regression. As the name suggests, SURE is an unbiased estimate of the out of sample prediction risk under an assumption of Gaussian errors. A tractable expression is not always available, but when it is available, SURE is known to be a Rao--Blackwellized version of the cross validation loss, a connection pointed out by \citet{Efron2004}. Consequently, minimizing SURE provides a natural approach for selecting $\nu$, if prediction is the main modeling goal.  We note here that similar formulas for SURE for the lasso regression have been worked out by \citet{zou2007degrees} and \citet{tibshirani2012degrees} and for the horseshoe regression by \citet{BhadraJMLR}.

Starting from the formula for SURE defined by \citet[Equation 2.11]{Efron2004}:
\begin{align*}
    \mathrm{SURE} &= \lVert y - \tilde y \rVert_2^2 + 2\sigma^2\sum_{i=1}^{r} \frac{\partial \tilde y_i}{\partial y_i}, 
\end{align*}
where the first term is an estimate for the squared bias for prediction and the second term is the so called \emph{degrees of freedom,} providing an estimate of the variance. In this way, SURE also makes the bias--variance tradeoff explicit.

Based on the definition of $\mathrm{SURE}$, we give an expression for it in each of the two cases we consider: the $n$-means model and the linear regression model. 
\begin{theorem}\label{th:sure_orthog}
    Denote with $\Var(\gamma\mid \hat\gamma,\nu)$ the posterior variance-covariance matrix, from the normal likelihood model with a prior specification on $\gamma$ that depends on a parameter $\nu$, where $y=Z\gamma + \varepsilon; Z=UD$ from the SVD decomposition as previously described. Then:
    \begin{align*}
    \mathrm{SURE}(\nu) = & \lVert y-\tilde y\rVert_2^2 + 2\sum_{i=1}^r\sigma^2d_i^2\Var(\gamma\mid \hat\gamma,\nu)_{i,i}
\end{align*}
\end{theorem}
The proof can be found on Appendix~\ref{pf:sure_orthog}. Theorem \ref{th:sure_orthog} means that in the case of an orthogonal design matrix $X$, we can work directly with the principal components, and reduce our $p$-dimensional integrals to $p$ individual $1-$dimension integrals. In the multivariate case without an orthogonal design matrix, we get a similar result for $\mathrm{SURE}$, and state it in Theorem~\ref{th:sure}.
\begin{theorem}\label{th:sure}
Consider a general prior for $\beta$, which we denote by $\pi(\beta)$, and the normal model as in Equation \eqref{eq:multivariate_likelihood}, with $\Sigma = \sigma^2I$, then:
\begin{align*}
\mathrm{SURE} = & \lVert y-\tilde y\rVert_2^2 + 2\mathrm{tr}(\Var(X\beta\mid y)),
\end{align*}
where we denote with $\tilde y$ the prediction of $y$, with an estimated $\tilde \beta=\mathbb{E}[\beta\mid y]$.
\end{theorem}
The proof of Theorem \ref{th:sure} can be found in Appendix~\ref{pf:sure_multivariate}.
We emphasize the notation change for the prior of $\beta$, now expressed as $\pi(\beta)$ to indicate that this applies to any proper prior, and not only to the exponential power prior. Theorem~\ref{th:sure} is a generalization of Theorem~\ref{th:sure_orthog}, and does not restrict the dimensions of $X$. When applying Theorem~\ref{th:sure} to our setting, we use $\Var_\nu(X\beta\mid y)$, and minimize $\mathrm{SURE}$ as a function of $\nu$, since $\tilde y$ is also a function of $\nu$ using the estimated $\tilde\beta = \mathbb{E}_\nu[\beta\mid y]$. As noted, SURE is an explicit numeric measure of the bias-variance trade-off, where the first term is a measure of the squared bias and the second of the variance.

With this in mind, it can be seen from the bias and variance terms that $\nu$ acts as an explicit parameter to control the bias-variance trade-off. When $\nu\to 0$, we have least squares regression and the bias is small, but the variance is large. When $\nu\to \infty$, the estimate becomes intercept only, which has zero variance but a potentially large bias. While these properties of $\nu$ are well known, what is not always available is a formulation of SURE as a function of $\nu$, which can then be passed on to a one-dimensional numerical optimizer to proceed via grid search. We have closed this gap through the expressions of desired fitted quantities.

\subsection{Tuning \texorpdfstring{$\nu$}{nu} by Minimizing SURE}

To estimate $\mathbb{E}_{\nu^*}[\beta\mid y]$ for $\nu^*$ the value that minimizes $\mathrm{SURE}$, we make use of Corollary~\ref{cor:simpler_formulas} and Theorem~\ref{th:sure}. 

First, simulate $T_{j,i}\overset{iid}{\sim} PS^+(\alpha,1/2)$ a polynomially-tilted stable random variable, for $j=1,\dots,M;\; i=1,\dots,p$, using the method described in \citet{devroye2009polynomiallytiltedStable}. Let $\mathbf{T}_j=(T_{j,1},\dots,T_{j,p})$ a $p$-dimensional vector. Define: $\Lambda_{\mathbf{T}_j}=\text{diag}(T_{j,1},\dots,T_{j,p})$, and compute $A_{\mathbf{T}_j}=X\Lambda_{\mathbf{T}_j}X^T$. Compute: $\log(p_\nu(y\mid \mathbf{T}_j)), {\mathbb{E}[X\beta\mid y,\mathbf{T}_j]}, \Var(X\beta\mid y, \mathbf{T}_j)$, using respectively the formula from Theorem~\ref{th:multivariate_case}, and from Corollary~\ref{cor:simpler_formulas}. For these we only need to compute the inverse of $(A_{\mathbf{T}_j} + \nu^{-1}\Sigma)$ once per $\nu$ and per $j$, which dramatically speeds up our performance for large $p$, since this is an $n\times n$ matrix.

Next, since we are not concerned about the value of the marginal, we use the $log-exp-sum$ trick. Specifically, define: 
\begin{align*}
    \log(w^*_{j}) &=  \log(p_\nu(y\mid \mathbf{T}_{j})) -\max_{j=1,\dots,M}(\log(p_\nu(y\mid \mathbf{T}_{j}))),\\
    w_{j} & = w^{*}_j\left(\sum_{j=1}^M w_{j}^*\right)^{-1}.
\end{align*}
Using these weights, we estimate the posterior mean as: $\tilde y=\sum_{j=1}^M w_j\mathbb{E}[X\beta\mid y,\mathbf{T}_j]$. Similarly,  the posterior variance is estimated by:
\begin{align*}
    \widehat{\Var}_\nu(X\beta\mid y) =&  \sum_{j=1}^M w_{j} \left(\Var(X\beta\mid y,\mathbf{T}_j)\right. \\
    & \left.+ \mathbb{E}[X\beta\mid y,\mathbf{T}_j]\mathbb{E}[X\beta\mid y,\mathbf{T}_j]^T\right)\\
    & - \mathbb{E}[X\beta\mid y]\mathbb{E}[X\beta\mid y]^{T}.
\end{align*}
To ensure that these are the correct estimates, we present the following corollary of Theorem~\ref{th:multivariate_case}.
\begin{corollary}\label{cor:hats_are_right}
The estimators for the posterior mean and variance satisfy:
\begin{align*}
    \tilde y & = X\mathbb{E}[\beta\mid y]_\mathcal{T},\\
    \widehat{\Var}_\nu(X\beta\mid y) & = X(\mathbb{E}[\beta\beta^T\mid y]_\mathcal{T} - \mathbb{E}[\beta\mid y]_\mathcal{T}\mathbb{E}[\beta\mid y]^T_\mathcal{T})X^T
\end{align*}
\end{corollary}
The proof of this corollary can be found in Appendix~\ref{pf:hats_are_right}. Now that we have all the ingredients, we proceed to optimize over $\nu$ in the expression $\mathrm{SURE}=\lVert y-\tilde y\rVert^2_2 +2\widehat{\mathrm{Var}}_\nu(X\beta\mid y)$.

To ensure we are optimizing on a smooth surface of $\nu$, and to minimize performance time, we simulate the polynomially-tilted stable random variables only $M\times p$ times, instead of repeating it several times. Then, we perform a one-dimensional minimization of $\mathrm{SURE}(\nu)$ for this fixed set of simulated $T$s. The optimization surface is smooth, as all the functions are infinitely differentiable as functions of $\nu$, for positive values of $\nu$.

The implementation is available online \href{ https://github.com/loriaJ/Sure-tuned_BridgeRegression}{https://github.com/loriaJ/Sure-tuned_BridgeRegression}, and we include a small example of how it can be executed. 

\subsection{Computational Time Complexity}\label{subsec:compcomplexity}
Computing the symmetric matrix $A_{\mathbf{T}}$, defined in Corollary~\ref{cor:simpler_formulas}, costs $\mathcal{O}(n^2p)$, as it only requires simulating the stable $\mathbf{T}_j$ random variables which costs $\mathcal{O}(1)$ \citep{devroye2009polynomiallytiltedStable}, followed by matrix multiplications. Next, the inverse of the $n\times n$ matrix $V_{\mathbf{T},\nu}$, defined in Theorem~\ref{th:multivariate_case}, costs $\mathcal{O}(n^3)$. The rest of the required computations have lower complexity, giving an overall complexity of $\mathcal{O}(n^2p)$ when $p>n$ for the proposed procedure, which is linear in $p$ for a given $n$. This favorable scaling in $p$ is verified via simulations in Supplementary Section~\ref{subsec:var_n_p}.
\section{Numerical Experiments}\label{sec:Results}

To measure the performance of our method, we compare it to the fully Bayesian method proposed by \citet{PolsonScottBayesBridge} (labeled ``BayesBridge'') as well as with an EM method by the same authors, choosing $\nu$ via cross validation (labeled ``Cross-validation''). We compare the following: (1) prediction error as measured by out-of-sample sum of squared errors (SSE), and (2) running time. For this, we simulate data sets $X,y$. Taking $X$ as a multivariate  normal with mean vector equal to zero, and with an equicorrelated covariance matrix with diagonal equal to one, and off-diagonal entries equal to $\rho$. We report the results for $\rho=0.9$ here, and in
the Supplementary Section~\ref{subsec:low_cors}, we report additional results for lower correlations ($\rho=0.1,0.5$) in the design matrix. We simulate $y$ using Equation~\eqref{eq:multivariate_likelihood}, setting $\Sigma$ as the identity matrix. We do this exercise for $p=1000,n=100$, and using $\beta$ with ten signals which equal ten, and the rest equal to zero, plus a normal noise with standard deviation of $0.1$. 

As a note, \citet{PolsonScottBayesBridge} propose two MCMC methods, which they call ``stable'' and ``triangular''. They both sample from a fully Bayesian model with the same marginal, but differ in the latent representation used. In our initial simulations the latter performed a lot worse in terms of SSE than the former, so we do not include comparisons with it.

\begin{figure*}[!htb]
   \centering
   \includegraphics[height=7cm,width=15cm]{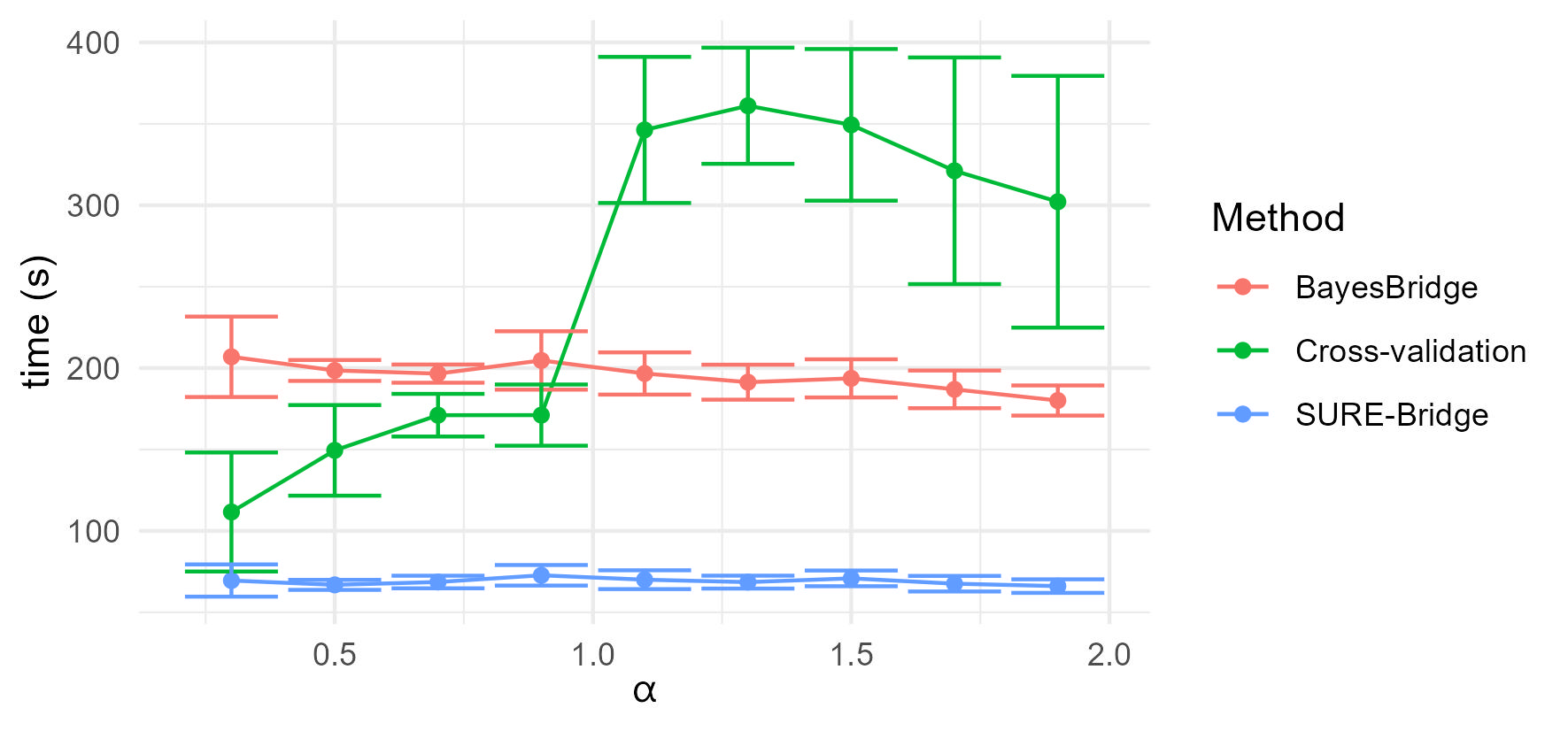}
   \caption{Comparison of average running time (s) $\pm$ SD by method, when changing the $\alpha$ parameter. Using  $n=100$, $p=1000$, in design matrices generated using $\rho = 0.9$.}
   \label{fig:line_plot_n_100_p_1000}
\end{figure*}

Figure \ref{fig:line_plot_n_100_p_1000} shows the comparison of average running time in seconds for the case of $p=1000,\rho=0.9,n=100$. Remarkably, our proposed method (termed ``SURE-Bridge'') runs in about one fourth of the time of BayesBridge, and is in general more computationally efficient than cross validation. The running times are non-uniform for the cross validation method and we comment further on this in Supplementary Section~\ref{subsec:cross validation}. 
Furthermore, the error bars indicate less variability around the mean running time compared to iterative approaches. We remark here that cross validation was parallelized across different folds, which means that hardware specification then enters the picture through the number of available processors, and complicates the comparison of the raw running times. However, for the SURE-Bridge method we did not parallelize it. Potentials for further speed up also exists for the proposed SURE-Bridge approach, for example, by farming out the vectorized vanilla Monte Carlo  calculations to a graphics processing unit or GPU. We have refrained from these engineering experiments in current work, and our figures paint an accurate picture of the raw running times for a single processor (possibly multi-threaded) machine without an explicit attempt at parallelization for the proposed method.

\begin{table*}[ht]
\centering
\caption{Average SSE (SD) by method in one hundred out of sample simulated datasets, by $\alpha$. Using $n=100$, $p=1000$, in a design matrix generated with  $\rho=0.9$.}
\label{tab:table_n_100_p_1000_cor_09}
\begin{tabular}{rllll}
  \hline
$\alpha$ & SURE & SURE-Bridge & BayesBridge & cross validation \\ 
  \hline
  0.30 & 198.89 (0.30) & 199.07 (28.04) & 170.82 (25.55) & 197.81 (47.95) \\ 
  0.50 & 198.76 (0.26) & 199.79 (29.61) & 191.16 (28.45) & 197.48 (27.15) \\ 
  0.70 & 198.78 (0.23) & 196.30 (30.80) & 196.08 (30.58) & 200.92 (31.23) \\ 
  0.90 & 198.79 (0.17) & 199.99 (26.44) & 199.89 (26.42) & 216.00 (32.4) \\ 
  1.10 & 198.83 (0.16) & 195.54 (29.63) & 195.50 (29.59) & 195.85 (29.21) \\ 
  1.30 & 198.87 (0.17) & 197.63 (27.70) & 197.62 (27.69) & 202.91 (49.49) \\ 
  1.50 & 198.88 (0.17) & 196.87 (28.43) & 196.90 (28.42) & 213.69 (94.49) \\ 
  1.70 & 198.89 (0.17) & 197.98 (24.06) & 198.03 (24.05) & 456.77 (842.21) \\ 
  1.90 & 198.90 (0.17) & 197.27 (30.98) & 197.34 (31.00) & 962.52 (1971.39) \\ 
   \hline
\end{tabular}
\end{table*}

The improved computational speed for SURE-Bridge is only meaningful if there is not a considerable price to be paid so far as the statistical performance is concerned. In Table~\ref{tab:table_n_100_p_1000_cor_09}, we show the SSE and their standard deviation (SD) for the methods under consideration, and the estimated SURE at the optimum $\nu=\nu^*$. Using $n=100$ and $p=1000$, with a design matrix produced using $\rho=0.9$, BayesBridge and SURE-Bridge have similar statistical performances over all the values of $\alpha$, while the cross validation approach has similar statistical performance except for $\alpha$ greater than $1.5$, where it performs poorly. Furthermore, SURE mostly falls within one standard deviation of the SSE for BayesBridge and SURE-Bridge.

To summarize the numerical results, then, our claimed achievement in this paper is not a better statistical estimator per se. Instead, the innovation lies in achieving competitive statistical performance, at a fraction of the running time for other methods, thereby facilitating the deployment of Bridge models at far larger problem dimensions. This is possible because we are able to bypass the iterative routines needed for the other methods through our closed form calculations and vanilla Monte Carlo approaches.

We present additional numerical results in Supplementary Section~\ref{subsec:var_n_p} to understand the scaling of the computational times with changes in $n$ and $p$ for all methods, where the advantage of our method stands out with larger $p$ compared to other approaches. We assess robustness to two departures from modeling assumptions: namely normality and independence of the error terms in Supplementary Section~\ref{subsec:devs}, that shows our method to be relatively robust to modeling violations. Further, we compare the efficiency of the parameter estimates in Supplementary Section~\ref{subsec:efficiency} for all methods, and find our method performs similarly to BayesBridge in signal recovery, although it is tuned to minimize out of sample prediction error, which is a different modeling goal compared to feature selection.

\section{Prediction of Photosynthetic Capacity with Spectroscopic Measurements}\label{sec:real_data}

To predict photosynthesis in plants \citet{MeachamHensoldMontesWuetal} use spectral measurements. They consider the photosynthetic capacity for their measured leaves by the maximum electron transport rate ($\mathcal{J}_{\max}$). Measuring the electron transport rate is expensive and time intensive. This prompted the research by \citet{MeachamHensoldMontesWuetal} to use spectral measurements of the leaves to be able to predict it, as they report that this is much less expensive. For this task, they collected $n=94$ observations and $p=2156$ covariates.


We consider this variable as the response ($y$). Using the leaf reflectances as the predictors ($X$). We repeat ten times splitting the data into equal-sized parts. We fit all methods in the training split and measure the prediction error as the SSE for the testing split. For each split we obtain an SSE per-method and an estimate of $\mathrm{SURE}$ from our method.

\begin{figure*}[!htb]
   \centering
   \includegraphics[height=7cm,width=15cm]{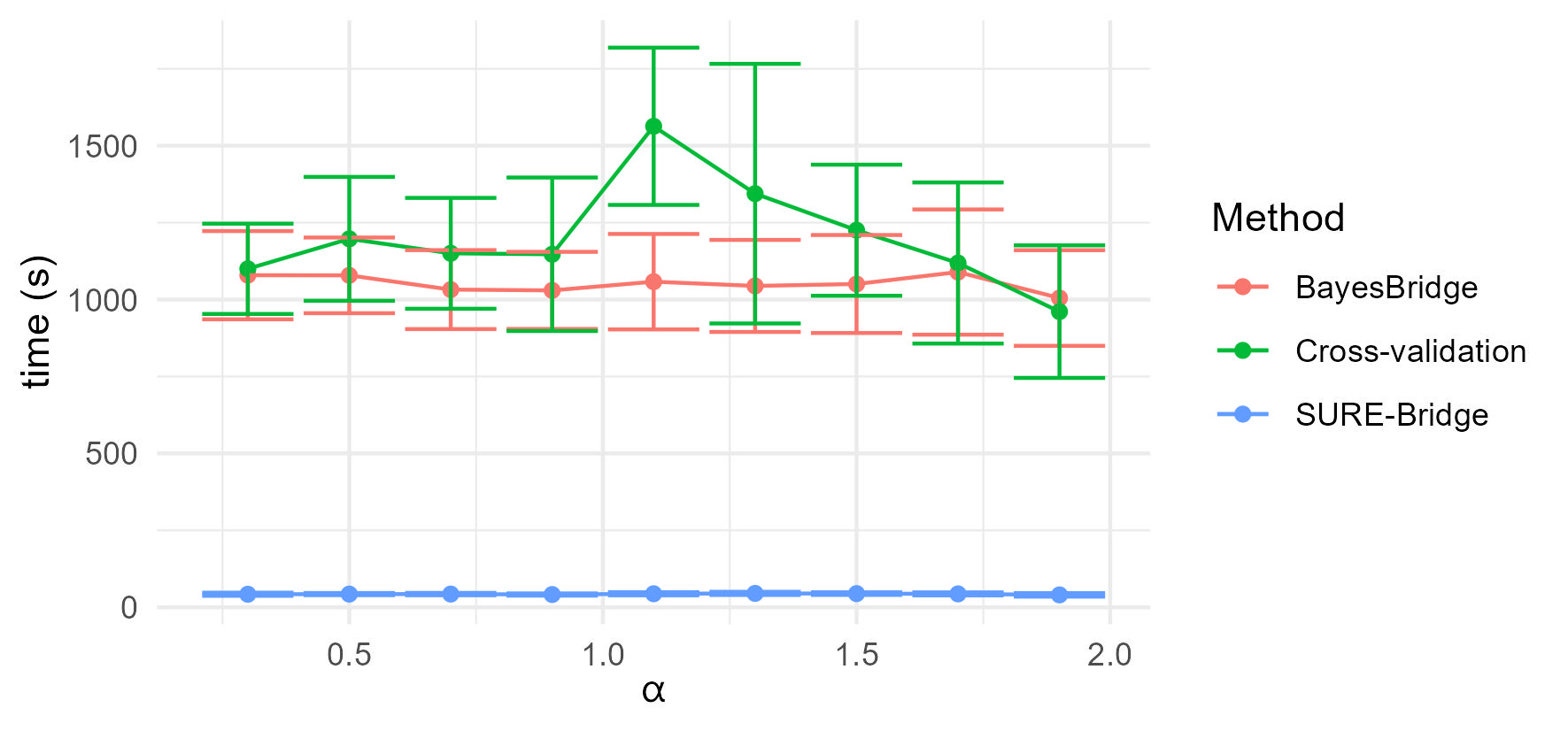}
   \caption{Comparison of running time (s) by method, for photosynthetic capacity data with $p=2156$ and $n=94$, based on ten splits}
   \label{fig:time_plot_Jmax}
\end{figure*}

\begin{table*}[ht]
\centering
\begin{tabular}{rllll}
  \hline
$\alpha$ & SURE & SURE-Bridge & BayesBridge & cross validation \\ 
  \hline
  0.30 & 32.60 (4.05) & 34.64 (6.07) & 34.10 (5.23) & 37.97 (8.08) \\ 
  0.50 & 32.56 (4.09) & 34.71 (6.11) & 34.04 (5.24) & 40.55 (11.45) \\ 
  0.70 & 32.56 (4.09) & 34.72 (6.12) & 33.83 (5.32) & 42.12 (9.60) \\ 
  0.90 & 32.56 (4.10) & 34.73 (6.12) & 33.77 (5.37) & 44.20 (14.41) \\ 
  1.10 & 32.56 (4.10) & 34.73 (6.13) & 33.70 (5.31) & 36.57 (8.07) \\ 
  1.30 & 32.56 (4.10) & 34.73 (6.12) & 33.75 (5.54) & 36.96 (8.64) \\ 
  1.50 & 32.56 (4.10) & 34.73 (6.13) & 33.71 (5.43) & 39.61 (13.23) \\ 
  1.70 & 32.56 (4.10) & 34.73 (6.13) & 33.71 (5.60) & 39.25 (11.32) \\ 
  1.90 & 32.56 (4.10) & 34.73 (6.13) & 33.71 (5.63) & 39.47 (10.04) \\ 
   \hline
\end{tabular}
\caption{Average prediction SSE (SD) by method in ten partitions of photosynthetic capacity data with $p=2156$ and $n=94$}
\label{tab:table_Jmax_sse}
\end{table*}

In Figure~\ref{fig:time_plot_Jmax} we display the time comparison between our method and the BayesBridge method \citep{PolsonScottBayesBridge} for the variable $\mathcal{J}_{\max}$. This only refers to the training portion. In this figure, it is clear how our method runs in about a tenth of the time of the competing methods. 

Furthermore, our results show that statistical accuracy is similar in all methods, see Table~\ref{tab:table_Jmax_sse}. In this table, we also show the estimated SURE, and how it is within one standard deviation of the SSE.

\section{Conclusion}\label{sec:Conclusion}

We have proposed a new \emph{non-iterative} approach for fitting the Bridge regression model, by selecting the tuning parameter through a one-dimensional numerical minimization of SURE. Once the tuning parameter is selected, using the latent Gaussian representation of the exponential power distribution yields the desired posterior moments in a tractable form that may be evaluated using vectorized vanilla Monte Carlo routines with well-behaved variances for the desired quantities. The construction of the Bridge regression estimate then follows from the equivalence between the Bayesian maximum a posteriori estimate and the penalized optimization problem stated at the very outset of this paper.  Consequently, our approach is non-iterative, yielding substantial computational gains over both fully Bayesian MCMC approaches as well as EM or coordinate descent algorithms for finding the maximum a posteriori estimate.

Throughout our calculations in this paper, we have assumed i.i.d. Gaussian error for the errors with known variance $\sigma^2$. If this condition is violated, SURE is not necessarily an unbiased estimation of the out of sample prediction risk \citep{stein1981estimation}. For our simulations, we worked with the true known $\sigma^2$ and for the spectroscopic data analysis we standardized both the predictors and responses and assumed i.i.d. standard normal error terms. The original paper by \citet{stein1981estimation} outlines a strategy for dealing with the unknown error variance case and shows it is still possible to construct an unbiased estimate of the prediction risk in the $n$-means case. However, in linear regression models under heteroskedastic normal or possibly non-normal errors this problem is still open, with some recent progress by \citet{xie2012sure}. Empirical and theoretical extensions of the techniques developed in this paper to potential violations of the modeling assumptions should be a promising direction for future works.

\section*{Supplementary Material}
The Supplementary Material contains additional simulation results. Computer code publicly is available from github at: \href{ https://github.com/loriaJ/Sure-tuned_BridgeRegression}{https://github.com/loriaJ/Sure-tuned_BridgeRegression}

\section*{Acknowledgements}
Bhadra is supported by Grant No. DMS-2014371 by the U.S. National Science Foundation.

\onecolumn

\pagestyle{plain}

\appendix
\section{Proofs}

\subsection{Proof of Lemma \ref{le:equivalence_between_models_n_means}}\label{pf:equivalence_between_models_n_means}
Using Equation~\eqref{eq:StableNormalMixture}, we have that the exponential power prior satisfies:
\begin{align*}
 p_\nu(\beta_i)\propto & \exp(-2^\alpha\nu^{\alpha}\lvert\beta_i\rvert^\alpha)\\
 = & \int_0^\infty p_\nu(\beta_i\mid t_i)t_i^{-1/2}p_L(t_i)dt_i,\\
 \propto & \int_0^\infty p_\nu(\beta_i\mid t_i)p_T(t_i)dt_i,
\end{align*}
where $p_\nu(\beta_i\mid t_i) = \exp(-\beta_i^2t_i/(2\nu)) (2\pi \nu/t_i)^{-1/2}$, and $p_T$ corresponds to the density of $T_i\overset{iid}{\sim}PS^+(\alpha/2,1/2)$, which by definition is proportional to $t_i^{-1/2}p_L(t_i)$. This completes the proof.

\subsection{Proof of Theorem \ref{th:n_means_moments}}\label{pf:n_means_moments}
For this proof we omit the sub-index $i$.

From Lemma~\ref{le:equivalence_between_models_n_means}, we integrate out $\beta$, and obtain $y\mid T \sim \mathcal{N}(0,\sigma^2 + \nu T^{-1})$, a property of the normal hierarchical model. This means that the marginal is given by:
\begin{align*}
    m_\nu(y)=\int_0^\infty p(y\mid t)p_T(t)dt,
\end{align*}
where $p_T(t)$ is the density of $T\sim PS^+(\alpha/2,1/2)$. Now, we show that it is bounded, as follows:
\begin{align*}
    m_\nu(y) & = \int_{\mathbb{R}^+}\frac{\exp(-(y^2/2)(\sigma^2+\nu t^{-1})^{-1})}{(2\pi)^{1/2}(\sigma^2+\nu t^{-1})^{1/2}}p_T(t)dt\\
    &< \frac{1}{(2\pi)^{1/2}}\int_{\mathbb{R}^+}\left(\frac{t}{\sigma^2t+\nu}\right)^{1/2}p_T(t)dt\\
    & < (2\pi\sigma^2)^{-1/2},
\end{align*}
where the first inequality follows by upper bounding the exponential term by one, and the second inequality follows from: $\sigma^2t(\sigma^2t+\nu)^{-1}<1$, and $\int_{0}^\infty p_T(t)dt=1$.

Now, for the posterior moment, we use iterated expectations as follows:
\begin{align*}
    \mathbb{E}[\beta\mid y] 
    = & \mathbb{E}[ \mathbb{E}[\beta\mid y,t]\mid y] \\
    = & \mathbb{E}[\nu y(\sigma^2T+\nu)^{-1} \mid y] \\
    = & \frac{1}{m_\nu(y)}\int_{0}^\infty \frac{\nu y}{\sigma^2t+\nu}p(y\mid t)p_T(t)dt,
\end{align*}
where on the second line we use that: $\beta\mid y,T\sim \mathcal{N}(\nu y(\sigma^2T+\nu)^{-1},\sigma^2\nu(\sigma^2T+\nu)^{-1})$, by Lemma~\ref{le:equivalence_between_models_n_means} and the normal hierarchical model, and the third line follows by Bayes' rule. Similarly, for the second posterior moment:
\begin{align*}
    \mathbb{E}[\beta^2\mid y] 
    = & \mathbb{E}[ \mathbb{E}[\beta^2\mid y,T]\mid y] \\
    = & \mathbb{E}[ \sigma^2\nu(\sigma^2 T + \nu)^{-1} + (\nu y(\sigma^2T+\nu)^{-1})^2 \mid y] \\
    = & \frac{1}{m_\nu(y)}\int_{0}^\infty \left(\frac{\sigma^2\nu}{\sigma^2 t + \nu} + \left(\frac{\nu y}{\sigma^2t+\nu}\right)^2\right)p(y\mid t)p_T(t)dt.
\end{align*}

\subsection{Proof of Lemma~\ref{le:marginal_bounded}}\label{pf:bounded_marginal}

Define the positive constant:
\begin{align*}
    C_{(\sigma^2,\nu)} = (2\pi\sigma^2)^{-1/2}\int_{0}^{\infty}\left(\frac{t}{ t+\nu(\sigma^2)^{-1}}\right)^{1/2}p_T(t)dt.
\end{align*}
Since the density $p_T$ is strictly positive the constant must be positive. It is finite since $t/(t+a)<1$ for all $a>0$. Now, by definition of the marginal:
\begin{align*}
    m_\nu(y) & = (2\pi)^{-1/2}\int_{0}^\infty \exp\left(-y^2\frac{t}{2(\sigma^2t+\nu)}\right)\left(\sigma^2+\nu t^{-1}\right)^{-1/2}p_T(t)dt,\\
    & = \exp\left(\frac{-y^2}{2\sigma^2}\right)\int_{0}^\infty\exp\left(+y^2\frac{\nu}{2\sigma^2(\sigma^2t+\nu)}\right)(2\pi)^{-1/2}\left(\sigma^2+\nu t^{-1}\right)^{-1/2}p_T(t)dt,\\
    & > (2\pi)^{-1/2}\exp\left(\frac{-y^2}{2\sigma^2}\right)\int_0^\infty\left(\sigma^2+\nu t^{-1}\right)^{-1/2}p_T(t)dt\\
    & = \exp\left(\frac{-y^2}{2\sigma^2}\right) C_{(\sigma^2,\nu)},
\end{align*}
where the second line follows by using, inside the exponential, that $t/(t+b) = 1 - b/(t+b)$, and appropriately multiplying by $-y^2/(2\sigma^2)$.
The last line follows by using the definition of $C_{(\sigma^2,\nu)}$ and that an exponential of a positive value is greater than 1.

\subsection{Proof of Theorem~\ref{th:finite_var}}\label{pf:th_finite_var}

Again, we omit the $i$. For simplicity call: $m_j=p(y\mid T_j)$, and $\Bar{m}= (m_1+\dots+m_J)/J$. Note that $\mathbb{E}[\Bar{m}\mid y] = \mathbb{E}[m_j\mid y]=m_\nu(y)$. Which proves that $m_\nu(y)_\mathcal{T}$ is an unbiased estimate. Next, let $s_j = \nu y(\sigma^2T_j+\nu)^{-1}p(y\mid T_j)$, $\Bar{s} = (s_1+\dots+s_J)/J$. 
By definition, we have that: 
\begin{align*}
    \mathbb{E}[\Bar{s}\mid y] & = \mathbb{E}[s_j\mid y]\\
    &= m_\nu(y)\mathbb{E}[\mathbb{E}[\beta\mid y,T]\mid y]\\
    &= m_\nu(y)\mathbb{E}[\beta\mid y],
\end{align*}
using Bayes' rule and iterated expectations. For what follows, we define $\mu_s = m_\nu(y)\mathbb{E}[\beta\mid y]$. Next, using independence of the $T_j$, we have that:
\begin{align*}
    \Var[m_\nu(y)_\mathcal{T}\mid y] & = \frac{1}{J^2}\sum_{j=1}^J \Var(m_j\mid y)\\
    &= J^{-1}\Var(m_j\mid y).
\end{align*}
Then, we have:
\begin{align*}
   \Var(m_j\mid y) &< \mathbb{E}[m_j^2\mid y]\\
   &=\int_{0}^\infty p(y\mid t)^2p_T(t)dt\\
   &= (2\pi)^{-1}\int_{0}^\infty \exp[-y^2(\sigma^2 + \nu t^{-1})] (\sigma^2 + \nu t^{-1})^{-1}p_T(t)dt\\
   & < (2\pi\sigma^2)^{-1}.
\end{align*}
The first inequality follows from the usual formula for the variance, the second equality from the definition of a normal random variable, and the second inequality follows since $\exp(-y^2(\sigma^2+\nu t^{-1})) <1$, and since $(\sigma^2+\nu t^{-1})^{-1}=t/(\sigma^2t+\nu)<1/\sigma^2$. This finishes the first part of the proof.

Now, for the second part: 
\begin{align*}
    \Var(s_j\mid y) &\leq \mathbb{E}[s_j^2\mid y],\\
    & =  \int_{0}^\infty \nu^2 y^2p(y\mid t)^2(\sigma^2 t + \nu)^{-2}p_T(t)dt,\\
    & < y^2 \int_{0}^\infty p(y\mid t)^2 p_T(t)dt,\\
    & <  (2\pi\sigma^2)^{-1} y^2,
\end{align*}
where the first inequality follows by the variance formula: $\Var(X)=\mathbb{E}(X^2)-\mathbb{E}(X)^2$, the second inequality follows from the fact that $\nu/(\nu + a)< 1$, for $a>0$, and the last inequality follows from the proof of finite variance of the marginal. This means that for large enough $J$, $\sqrt{J}(\Bar{s}-\mu_s)\sim \mathcal{N}(0,\sigma_s^2)$, and $\sqrt{J}(\Bar{m}-\mu_m)\sim \mathcal{N}(0,\sigma_m^2)$.

Further, define $\sigma_{sm} =  \mathrm{Cov}(s_j,m_j\mid y)$, and we have: $\lvert \sigma_{sm}\rvert < \lvert y\rvert (2\pi\sigma^2)^{-1}$. Using the delta method we have that:
\begin{equation*}
    \sqrt{J}\left(\frac{\Bar{s}}{\Bar{m}} - \frac{\mu_s}{\mu_m}\right)\sim\mathcal{N}(0, \nabla h(\mu_s,\mu_m)^T \Sigma \nabla h(\mu_s,\mu_n)),
\end{equation*}
where $h(a,b) = a/b$, and 
\begin{equation*}
    \Sigma = \begin{bmatrix}
    \sigma_s^2 & \sigma_{sm}\\
    \sigma_{sm} & \sigma_m^2
    \end{bmatrix}.
\end{equation*}
First note that this means that our estimate is asymptotically unbiased, as $\mu_s/\mu_m=\mathbb{E}[\beta\mid y]$. Next, since $\nabla h(a,b) = (1/b,-a/b^2)$, the variance term above becomes: 
\begin{align*}
    \nabla h(\mu_s,\mu_m)^T \Sigma \nabla h(\mu_s,\mu_n) &= \sigma^2_s\mu_{m}^{-2}  - 2\sigma_{sm}\mu_s\mu_m^{-3} + \sigma^2_m\mu_s^2\mu_{m}^{-4}.
\end{align*}
Plugging in the values for $\mu_s,\mu_m$ and using the inequalities we derived before, this becomes: 
\begin{align*}
\Var(\mathbb{E}[\beta\mid y]_\mathcal{T}\mid y) &= m_\nu(y)^{-2}(\sigma_s^2 - 2 \sigma_{sm}\tilde{\beta}+\tilde{\beta}^2)\\
 &<(2\pi\sigma^2)^{-1}m_\nu(y)^{-2}(\lvert y\rvert+ \tilde{\beta})^2,
\end{align*}
which gives the second bound we stated.

Next, denote $v_j = p(y\mid t_j)(\sigma^2\nu(\sigma^2t_j+\nu)^{-1} + y^2\nu^2(\sigma^2t_j+\nu)^{-2})$, and $\Bar{v}=(v_1+\dots + v_J)/J$. Then:
\begin{align*}
    \Var(v_j\mid y)& \leq \mathbb{E}[v_j^2\mid y]\\
    &=  \int_0^\infty p(y\mid t)^2\left(\sigma^2\nu(\sigma^2t+\nu)^{-1} + y^2\nu^2(\sigma^2t+\nu)^{-2}\right)^2p_T(t)dt,\\
    &< \int_0^\infty \left( \sigma^2 + y^2\right)^2p(y\mid t)^2 p_T(t)dt\\
    & < (2\pi\sigma^2)^{-1}(\sigma^2 + y^2)^2 \int_0^\infty p_T(t)dt,
\end{align*}
where the second inequality follows since for any positive $a$: $\nu/(a+\nu) < 1,\nu^2/(\nu + a)^2<1$, and since $g(x)=x^2$ is a monotone function for $x>0$. The third inequality follows by using the first part of the proof. Similarly, $\mathrm{Cov}(v_j,m_j\mid y) < (2\pi\sigma^2)^{-1}(\sigma^2 +y^2)$. The rest of the proof follows as the second part, by replacing the $s$ sub-indices with $v$ sub-indices and we obtain:
\begin{align*}
    \Var(\mathbb{E}[\beta^2\mid y]_\mathcal{T} \mid y)&= (2\pi\sigma^2)^{-1}m_\nu(y)^{-2}((\sigma^2 + y^2)^2 -2(\sigma^2+y^2)\tilde{\beta}^{(2)} + (\tilde{\beta}^{(2)})^2)\\
    & = (2\pi\sigma)^{-1}m_\nu(y)^{-2}(\sigma^2 + y^2-\tilde{\beta}^{(2)})^2.
\end{align*}

\subsection{Proof of Lemma \ref{le:equivalence_between_models}} \label{pf:equivalence_between_models}
Using Equation~\eqref{eq:StableNormalMixture}, we have that the exponential power prior satisfies:
\begin{align*}
 p_\nu(\beta_i)\propto & \exp(-\nu\lvert\beta_i\rvert^\alpha)\\
 = & \int_0^\infty p_\nu(\beta_i\mid t_i)t_i^{-1/2}p_L(t_i)dt_i,\\
 = & \int_0^\infty p_\nu(\beta_i\mid t_i)p_T(t_i)dt_i,
\end{align*}
where $\beta_i\mid t_i \overset{indep.}{\sim}\mathcal{N}(0,\nu t_i^{-1})$, and the density $p_T$ corresponds to $T\sim PS^+(\alpha/2,1/2)$, which completes the proof as the likelihood does not change by adding the information of $T$.

\subsection{Proof of Theorem \ref{th:multivariate_case}} \label{pf:multivariate_lemma}
Using Lemma~\ref{le:equivalence_between_models}, we get: 
\begin{align*}
    m_\nu(y) = & \int_{\mathbb{R}^p}\int_{(\mathbb{R}^{+})^p} p(y\mid \beta,\mathbf{t})\prod_{i=1}^pp(\beta_i\mid t_i)p_T(t_i)d\beta_i dt_i\\
    = &\int_{(\mathbb{R}^+)^p} \int_{\mathbb{R}^p}p(y\mid \beta,\mathbf{t})\prod_{i=1}^p p(\beta_i, t_i) d\beta dt_i, \\
    = & \int_{(\mathbb{R}^+)^p}p(y\mid T) \int_{(\mathbb{R}^+)^p}p(\beta\mid y,\mathbf{t}) d\beta\prod_{i=1}^p p_T(t_i) dt_i,\\
    = & \int_{(\mathbb{R}^+)^p}p(y\mid \mathbf{t})\prod_{i=1}^p p_T(t_i) dt_i,
\end{align*}
where $y\mid \mathbf{t}\sim\mathcal{N}(0,\nu V_{\mathbf{t},\nu}), V_{\mathbf{t},\nu} = X\Lambda_\mathbf{t}X^T+\nu^{-1} \Sigma$, by using the fact that:
\begin{align*}
 (\beta,y)^T\mid \mathbf{t} \sim \mathcal{N}\left((0_n,0_p)^T,
 \begin{bmatrix}
 \Lambda_\mathbf{t} & \Lambda_\mathbf{t}X^T\\
 X\Lambda_\mathbf{t} & \nu V_{\mathbf{t},\nu}
 \end{bmatrix} \right).
\end{align*}
We also need to prove that $m_\nu(y)$ is finite. For this we do: 
\begin{align*}
    m_\nu(y) & < (2\pi)^{-p/2}\int_{(\mathbb{R}^+)^p}\det(\nu V_{\mathbf{t},\nu})^{-1/2}\prod_{i=1}^pp_T(t_i)dt_i,\\
     & <(2\pi)^{-p/2}\det(\Sigma)^{-1/2}\int_{(\mathbb{R}^+)^p}\prod_{i=1}^pp_T(t_i)dt_i,
\end{align*}
where the first inequality follows since the exponential of a negative value is bounded by one, the second inequality since $\Sigma \preceq \nu V_{t,\nu}$, which means that $\det(\Sigma)< \det(\nu V_{t,\nu})$. Similarly,
\begin{align*}
    \mathbb{E}[\beta\mid y] = & \mathbb{E}[\mathbb{E}[\beta\mid y,T]\mid y] \\
    = & \frac{1}{m_{\nu}(y)}\int_{(\mathbb{R}^+)^p}\mathbb{E}[\beta\mid y,\mathbf{t}]p_\nu(y\mid \mathbf{t})\prod_{i=1}^p p_T(t_i)dt_i,
\end{align*}
by applying Lemma \ref{le:equivalence_between_models}. The expectation inside the integral can be computed using $\beta\mid y,\mathbf{t}\sim\mathcal{N}(\Lambda_\mathbf{t}X^TV_{\mathbf{t},\nu}^{-1}y,(\Lambda_{\mathbf{t}}^{-1} + X^T\Sigma X )^{-1})$, using the posterior normal formula. Now, for the second moment:
\begin{align*}
    \mathbb{E}[\beta\beta^T\mid y] & = \mathbb{E}[\mathbb{E}[\beta\beta^T\mid y,T]\mid y] \\
    & = \frac{1}{m_{\nu}(y)}\int_{(\mathbb{R}^+)^p}\mathbb{E}[\beta\beta^T\mid y,\mathbf{t}]p(y\mid \mathbf{t}) \prod_{i=1}^p p_T(t_i)dt_i.
\end{align*}

We comment on the invertibility of $V_{\mathbf{t},\nu}$, which was assumed throughout the proof. Consider $z\in\mathbb{R}^{n}\setminus\{\vec 0\}$, we name $w=X^Tz$, and $z^TX \Lambda_\mathbf{t} X^Tz + z^Tz=w^T\Lambda_\mathbf{t}w + z^Tz$, since $\Lambda_\mathbf{t}$ is a diagonal matrix with positive entries the first term is non-negative, and the second term is positive. This implies that $V_{\mathbf{t},\nu}$ is positive definite, as is required for a proper variance matrix; which concludes the proof.

\subsection{Proof of Corollary \ref{cor:simpler_formulas}}\label{pf:simpler_formulas}

The first equality follows from ${\mathbb{E}[X\beta\mid y,\mathbf{t}] = X\mathbb{E}[\beta\mid y,\mathbf{t}]}$, and using Theorem~\ref{th:multivariate_case}. The second equality similarly follows from Theorem~\ref{th:multivariate_case} and using the Woodbury matrix identity.

\subsection{Proof of Lemma \ref{le:bounded_marginal_vector}}\label{pf:bounded_marginal_vector}
Define:
\begin{align*}
    C_{(\Sigma,\nu,X)}& = \int_{(\mathbb{R}^+)^p}\det(\nu X\Lambda_\mathbf{t}X^T +\Sigma)^{-1/2}(2\pi)^{-p/2}\prod_{i=1}^pp_T(t_i)dt_i.
\end{align*}
Since the matrix inside the determinant is positive definite this determinant will be positive, which implies that $C_{(\Sigma,\nu,X)}>0$. This integral will be finite since $\nu V_{\mathbf{t},\nu}\succeq \Sigma$, which implies that $\det(\nu V_{\mathbf{t},\nu}) > \det(\Sigma)$. That is: $\det(\nu V_{\mathbf{t},\nu})^{-1/2}<\det(\Sigma)^{-1/2}$. By definition of the marginal:
\begin{align*}
    m_\nu(y) & = \int_{(\mathbb{R}^+)^p}p(y\mid \mathbf{t})\prod_{i=1}^pp_T(t_i)dt_i\\
    & = \exp(-y^T\Sigma^{-1}y/2)\int_{(\mathbb{R}^+)^p}\exp(B_y^T(\Lambda_{\mathbf{t}}^{-1} + X^T\Sigma^{-1}X)^{-1}B_y/2)\det(\nu V_{\mathbf{t},\nu})^{-1/2}\prod_{i=1}^pp_T(t_i)dt_i\\
    & > \exp(-y^T\Sigma^{-1}y/2)\int_{(\mathbb{R}^+)^p}
     \det(\nu V_{\mathbf{t},\nu})^{-1/2}
    \prod_{i=1}^pp_T(t_i)dt_i\\
    & = \exp(-y^T\Sigma^{-1}y/2)C_{(\Sigma,\nu,X)},
\end{align*}
where the second line follows by using Woodbury's formula and defining the $p$-dimensional real vector $B_y= X^T\Sigma^{-1}y$, the first inequality follows since the inverse of a positive definite matrix is positive definite and the exponential of a positive value will be greater than one.

\subsection{Proof of Theorem \ref{th:finite_var_vector}}\label{pf:finite_var_vector}
Define $m_j=p(y\mid \mathbf{t}^j)$, and $\Bar{m}=(m_1+\dots+m_J)/J$.  As a direct consequence of this, $\mathbb{E}[\Bar{m}]=\mathbb{E}[m_j]=m_\nu(y)$, which means $\Bar{m}$ is an unbiased estimator of $m_\nu(y)$. Next, let $s_j= p(y\mid \mathbf{t}^j)\Lambda_{\mathbf{t}^j}X^TV^{-1}_{\mathbf{t}^{j}}y$,  $\Bar{s}=(s_1+\dots+s_J)/J$,  
It follows that:
\begin{align*}
 \mathbb{E}[\Bar{s}] & = \mathbb{E}[s_j] \\
 & =\int_{(\mathbb{R}^+)^p}p(y\mid \mathbf{t})\Lambda_\mathbf{t}X^TV_{\mathbf{t},\nu}y\prod_{i=1}^pp_T(t_i)dt_i \\
 & = \tilde{\beta}m_\nu(y).
\end{align*}
For clarity of what follows, we denote the mean of $\Bar{s}$ by $\mu_s=\tilde{\beta}m_\nu(y)$. We proceed by proving the variance bounds for $m_j$ and $s_j$. The bound for $m_j$ will imply the first bound we state in the theorem, and the second bound we state will follow by using the delta method. For this we will need to know that: $\lVert \nu V_{\mathbf{t},\nu}\rVert_2 \leq \lVert \Sigma^{-1}\rVert$, and $\det(\nu V_{\mathbf{t},\nu})>\det(\Sigma)$, both of which follow by using the fact $\Sigma \preceq \nu V_{\mathbf{t},\nu}$. 

Now, the first bound:
\begin{align*}
    \Var(m_\nu(y)_\mathcal{T}\mid y) &= J^{-1}\Var(m_j)\\
    &< J^{-1}\mathbb{E}[m_j^2\mid y]\\
    & = \int_{(\mathbb{R}^+)^p}p_\nu(y\mid \mathbf{t})^2 \prod_{i=1}^p p_T(t_i)dt_i\\
    & < J^{-1}\int_{(\mathbb{R}^+)^p}(2\pi)^{-p} \det(V_{\mathbf{t},\nu}^{-1})\prod_{i=1}^p p_T(t_i)dt_i\\
    & \leq J^{-1}(2\pi)^{-p}\int_{(\mathbb{R}^+)^p} \det(\Sigma^{-1}) \prod_{i=1}^p p(t_i)dt_i\\
    &=  J^{-1}(2\pi)^{-p}\det(\Sigma^{-1}),
\end{align*}
where the first inequality is a well known property of the variance. Next, is the definition of expectation. Then, the inequality mentioned in the previous paragraph. Finally, since $p_T(t_i)$ are probability density functions which integrate to $1$. 

Now, we repeat this for the norm of $s_j$:
\begin{align*}
    \Var(\lVert s_j \rVert_2\mid y) &< \mathbb{E}[\lVert s_j\rVert_2^2 \mid y],\\
    &= \int_{(\mathbb{R}^+)^p}p(y\mid \mathbf{t})^2 \lVert \Lambda_\mathbf{t}X^TV_{\mathbf{t},\nu}^{-1}y\rVert_2^2\prod_{i=1}^p p_T(t_i)dt_i\\
    & < (2\pi)^{-p}\lVert y\rVert^2_2 \int_{(\mathbb{R}^+)^p}\det(\nu V_{\mathbf{t},\nu}^{-1})\lVert\Lambda_\mathbf{t}\rVert_2^2\lVert X^T\rVert_2^2\lVert V_{\mathbf{t},\nu}^{-1}\rVert_2^2 \prod_{i=1}^p p_T(t_i)dt_i\\
    &  < (2\pi)^{-p}\nu^2\det(\Sigma^{-1})\lVert\Sigma^{-1}\rVert_2^{2}\lVert y \rVert_2^2\lVert X^T\rVert_2^2 \int_{(\mathbb{R}^+)^p}\sum_{i=1}^p t_{i}^{-2}\prod_{i=1}^pp_T(t_i)dt_i\\
     & < \det(\Sigma^{-1})(2\pi)^{-p}\nu^2 pK_2\lVert\Sigma^{-1}\rVert_2^{2}\lVert y \rVert^2_2\lVert X^T\rVert^2_2.
\end{align*}

Similar to before, the first inequality follows by a known property of the variance, next is using the definition of expectation. The second inequality follows since the product of the norms is greater than the norm of the product. The third inequality is by using the inequalities of the first paragraph. Next is by using the definition of the Euclidean-norm. Finally, we have $K_2=\int_{\mathbb{R}^+}t^{-2}p_T(t)dt<\infty$, since the negative moments of $\alpha/2$-stable random variables are finite.
This means that the variance of both $\Bar{s}$ and $\Bar{m}$ are finite. By central limit theorem, for big enough $J$ we have:
\begin{align*}
 \sqrt{J}(\Bar{m}-m_\nu(y))&\sim\mathcal{N}(0,\sigma_m^2),  \\ 
 \sqrt{J}(\Bar{s} - \mu_s)&\sim\mathcal{N}(0,\Sigma_s).
\end{align*}
We need a bound on the norm of the covariance between $m_j$ and $s_j$:
\begin{align*}
 \lVert \mathrm{Cov}(m_j,s_j\mid y)\rVert_2 &= \left\lVert \int_{(\mathbb{R}^+)^p}p(y\mid \mathbf{t})^2 \Lambda_\mathbf{t}X^TV_{\mathbf{t},\nu}^{-1}y\prod_{i=1}^pp_T(t_i)dt_i   \right\rVert_2\\
 & < \int_{(\mathbb{R}^+)^p}p(y\mid \mathbf{t})^2 \lVert\Lambda_\mathbf{t}X^TV_{\mathbf{t},\nu}^{-1}y\rVert_2\prod_{i=1}^pp_T(t_i)dt_i\\
 & < (2\pi)^{-p}\det(\Sigma^{-1}) \int_{(\mathbb{R}^+)^p}\lVert\Lambda_\mathbf{t}X^TV_{\mathbf{t},\nu}^{-1} y\rVert_2\prod_{i=1}^pp_T(t_i)dt_i\\
 & < (2\pi)^{-p}\det(\Sigma^{-1})\nu\lVert X^T\rVert_2 \lVert \Sigma^{-1}\rVert_2 \lVert y \rVert_2\int_{(\mathbb{R}^+)^p} \sum_{i=1}^p \lVert t_i^{-1}\rVert_2\prod_{i=1}^pp_T(t_i)dt_i\\
 & = (2\pi)^{-p}\det(\Sigma^{-1}) \nu\lVert X^T\rVert_2 \lVert y\rVert_2\lVert \Sigma^{-1}\rVert_2 p K_1,
\end{align*}
where the first line is definition of the $p$-dimensional covariance, second line is using the fact that the norm of an integral is smaller than the integral of the norm, next is using the bounds established above for $p(y\mid \mathbf{t})$. The third inequality follows by the submultiplicative property of the norm and the triangle inequality on $\Lambda_\mathbf{t}$. The last line follows from independence of the $t_i$'s, and defining $K_1 = \int_{\mathbb{R}^+}t^{-1}p_T(t)dt$.

By definition of $\Bar{m},\Bar{s}$ we have the vector: $\mathrm{Cov}(\Bar{m},\Bar{s}\mid y) = J^{-1}\mathrm{Cov}(m_j,s_j\mid y)=J^{-1} \sigma_{s,m}$. As mentioned before, we use the delta method:
\begin{align*}
    \sqrt{J}\left(\frac{\Bar{s}}{\Bar{m}} - \frac{\mu_s}{\mu_m}\right)&\sim \mathcal{N}\left(0,\nabla g^T \Sigma^*g\right),\\
    \text{where: }\Sigma^* &=  \begin{pmatrix}
    \Sigma_s & \sigma_{s,m}^T\\
    \sigma_{s,m} & \sigma_m^2
    \end{pmatrix}\text{,}\\
    h(a,b) & = (a_1/b,\dots,a_p/b)\text{, and}\\
    g & = \nabla h(\mu_s,m_\nu(y)).
\end{align*}
This means that: 
\begin{align*}
    \nabla h(a,b)=\left(1/b,\dots,1/b,-\sum_{i=1}^p a_i/b^2\right),
\end{align*}
which implies that:
\begin{align*}
    \Var(\Bar{s}/\Bar{m}) = & J^{-1}\left(m_\nu(y)^{-2}\Sigma_s + \sigma^2_m \left(\sum_{i=1}^p \mu_{s,i}\right)^2m_\nu(y)^{-4}+2\sum_{i=1}^p\sigma_{s,m,i}\mu_{s,i}m_\nu(y)^{-3}\right).
\end{align*}
The rest of this bound follows by using Cauchy-Schwarz inequality and the bounds proved before:
\begin{align*}
 \Var(\lVert \mathbb{E}[\beta\mid y]_\mathcal{T}\rVert_2) & < J^{-1}m_\nu(y)^{-2}(\nu^2 pK_2\lVert\Sigma^{-1}\rVert_2^{2}\lVert y \rVert^2_2\lVert X^T\rVert^2_2 + \lVert\tilde\beta\rVert^2_2 + 2\nu\lVert X^T\rVert_2 \lVert y\rVert_2\lVert \Sigma^{-1}\rVert_2 p K_1 \lVert\tilde{\beta}\rVert_2).
\end{align*}

Now, we need to verify that $\mathbb{E}[\beta\beta^T\mid y]_\mathcal{T}$ will be asymptotically unbiased, and we also want to bound its variance. With this in mind, define:
\begin{align*}
    v_j = p(y\mid \mathbf{t}^j)\left(\nu\Lambda_{\mathbf{t}^j} - \nu\Lambda_{\mathbf{t}^j}X^TV_{\mathbf{t}^j,1}^{-1}X\Lambda_{\mathbf{t}^j}+\Lambda_{\mathbf{t}^j}X^T V_{\mathbf{t}^j,\nu}^{-1}yy^TV_{\mathbf{t}^j,\nu}^{-1} X\Lambda_{\mathbf{t}^j}\right),
\end{align*}
and $\Bar{v}=(v_1+\dots+v_J)/J$. Then we have,
\begin{align*}
    \mathbb{E}[\Bar{v}] &= \mathbb{E}[v_j]\\
    &= \int_{(\mathbb{R}^+)^p}p(y\mid \mathbf{t})\left(\nu\Lambda_{\mathbf{t}} - \nu\Lambda_{\mathbf{t}}X^TV_{\mathbf{t},1}^{-1}X\Lambda_{\mathbf{t}}+\Lambda_{\mathbf{t}}X^T V_{\mathbf{t},1}^{-1}yy^TV_{\mathbf{t},1}^{-1} X\Lambda_{\mathbf{t}}\right) \prod_{i=1}^p p_T(t_i)dt_i\\
    & = \int_{(\mathbb{R}^+)^p} (\Var(\beta\mid y,T)+ \mathbb{E}(\beta\mid y,\mathbf{t})\mathbb{E}(\beta^T\mid y,\mathbf{t}))p(y\mid \mathbf{t})\prod_{i=1}^p p_T(t_i)dt_i\\
    & = \int_{(\mathbb{R}^+)^p} \mathbb{E}[\beta\beta^T\mid y,\mathbf{t}]p(y\mid \mathbf{t})\prod_{i=1}^p p_T(t_i)dt_i\\
    & = m_\nu(y)\mathbb{E}[\mathbb{E}[\beta\beta^T\mid y,\mathbf{t}]\mid y]\\
    & = m_\nu(y)\mathbb{E}[\beta\beta^T\mid y],
\end{align*}
where the first equality follows from definition of $\Bar{v}$, the second one from definition of $v_j$. The third one, using the formulas derived in Theorem~\ref{th:multivariate_case}. The next equality by using the formula $\mathbb{E}[\beta\beta^T]= \Var(\beta) + \mathbb{E}[\beta]\mathbb{E}[\beta]^T$, finally by using Bayes' Theorem and the iterated expectations property. Similarly, we define ${\mu_v = m_\nu(y)\mathbb{E}[\beta\beta^T\mid y]}$.

Now, we proceed to prove the bound, as follows.
\begin{align*}
    \Var(\lVert v_j\rVert\mid y) &< \mathbb{E}[\lVert v_j\rVert^2 \mid y ]\\
    & = \int_{(\mathbb{R}^+)^p}p(y\mid \mathbf{t})^2 \left\lVert\nu\Lambda_{\mathbf{t}} - \nu\Lambda_{\mathbf{t}}X^TV_{\mathbf{t},\nu}^{-1}X\Lambda_{\mathbf{t}}+ \Lambda_{\mathbf{t}}X^T V_{\mathbf{t},\nu}^{-1}yy^TV_{\mathbf{t},\nu}^{-1} X^T\Lambda_{\mathbf{t}}\right\rVert_2^2\\
    & \hspace{3cm}\times\prod_{i=1}^p p_T(t_i)dt_i,\\
    & < (2\pi)^{-p}\det(\Sigma^{-1}) \int_{(\mathbb{R}^+)^p}\left(\left\lVert\nu\Lambda_{\mathbf{t}} - \nu\Lambda_{\mathbf{t}}X^TV_{\mathbf{t},\nu}^{-1}X\Lambda_{\mathbf{t}}\right\rVert_2 + \left\lVert\Lambda_{\mathbf{t}}X^T V_{\mathbf{t},\nu}^{-1}yy^TV_{\mathbf{t},\nu}^{-1} X\Lambda_{\mathbf{t}}\right\rVert_2\right)^2\\
    & \hspace{3cm}\times \prod_{i=1}^pp_T(t_i)dt_i\\
    &< (2\pi)^{-p}\det(\Sigma^{-1}) \int_{(\mathbb{R}^+)^p}\left(\left\lVert\nu\Lambda_{\mathbf{t}}\right\rVert_2+ \lVert \Lambda_\mathbf{t}\rVert_2^2\lVert V_{\mathbf{t},\nu}^{-1}\rVert_2^2\lVert y\rVert_2^2\lVert X\rVert^2_2  \right)^2\\
    & \hspace{3cm}\times\prod_{i=1}^p p(t_i)dt_i,\\
    & < (2\pi)^{-p}\det(\Sigma^{-1})\int_{(\mathbb{R}^+)^p}\left(\nu\left\lVert\Lambda_{\mathbf{t}}\right\rVert_2 +\nu^2 \lVert\Lambda_\mathbf{t}\rVert^2_2 \lVert\Sigma^{-1}\rVert^2_2\lVert X\rVert^2_2 \lVert y\rVert^2_2 \right)^2 \\
    & \hspace{3cm}\times\prod_{i=1}^p p_T(t_i)dt_i\\
    & <(2\pi)^{-p}\det(\Sigma^{-1})\int_{(\mathbb{R}^+)^p} \left\{\nu\left(\sum_{i=1}^p t_{i}^{-2}\right)^{1/2} +M\sum_{i=1}^pt_i^{-2} \right\}^2\\
    &\hspace{3cm}\times\prod_{i=1}^pp_T(t_i)dt_i\\
   & < (2\pi)^{-p}\det(\Sigma^{-1}) \int_{(\mathbb{R}^+)^p} \left(\sum_{i=1}^p\nu^2t_i^{-2} +2M\nu\sum_{i=1}^pt_i^{-2}\sum_{i=1}^pt_i^{-1} +M^2 \bigg(\sum_{i=1}^pt_i^{-4}+\sum_{i\neq j}t_i^{-2}t_j^{-2}\bigg)\right)\\
   & \hspace{3cm}\times \prod_{i=1}^pp_T(t_i)dt_i\\
   & = (2\pi)^{-p}\det(\Sigma^{-1})\left(p\nu^2K_2 + 2M\nu p K_3+ M\nu\frac{p(p-1)}{2}K_2K_1+M^2K_4+M^2p(p-1)K_2^2\right).
\end{align*}
The first line follows from the variance computation formula, the second line follows by definition of expectation, the second inequality uses triangle inequality and the inequality proved above for $p(y\mid \mathbf{t})$, the third inequality by the property that the norm of a product is smaller than the product of the norms and $\Lambda_\mathbf{t} -\Lambda_\mathbf{t}X^TV_{\mathbf{t},1}^{-1}X\Lambda_\mathbf{t} \preceq \Lambda_\mathbf{t}$, the fourth inequality since $V_{\mathbf{t},\nu}^{-1}\preceq \nu\Sigma^{-1}$, next one by using the definition of the norm and defining $M=\nu^2\lVert\Sigma^{-1}\rVert^2_2\lVert X\rVert^2_2 \lVert y\rVert^2_2$, and the last line follows since all the negative moments of $\alpha/2$-stable random variables are finite. This means that the variance of $v_j$ is finite.

Now, we need to compute a bound on the norm of the covariance between $v_j$ and $m_j$, we use the fact that:
\begin{align*}
    \lVert v_j\rVert_2 &< (2\pi)^{-p/2}\det(\Sigma^{-1})^{1/2} \left(\nu \left(\sum_{i=1}^pt_i^{-2}\right)^{1/2} + M\sum_{i=1}^pt_i^{-2}\right),
\end{align*}
which follows from the proof on the bound of the variance of $v_j$. Using this inequality we have that
\begin{align*}
    \lVert \mathrm{Cov}(v_j,m_j\mid y)\rVert_2 &<  \int_{(\mathbb{R}^+)^p} \lVert v_j\rVert_2 m_j\prod_{i=1}^pp_T(t_i)dt_i\\
    &< (2\pi)^{-p}\det(\Sigma^{-1})\int_{(\mathbb{R}^+)^p}\left\{\nu \left(\sum_{i=1}^pt_i^{-2}\right)^{1/2}+ M\sum_{i=1}^pt_i^{-2}\right\}\\
    &\hspace{2cm} \times \prod_{i=1}^pp_T(t_i)dt_i\\
    & < (2\pi)^{-p}\det(\Sigma^{-1})\int_{(\mathbb{R}^+)^p} \left(\nu \sum_{i=1}^pt_i^{-1}+M\sum_{i=1}^{p}\right)\\
    & \hspace{2cm}\times \prod_{i=1}^pp_T(t_i)dt_i\\
    & < (2\pi)^{-p}\det(\Sigma^{-1}) (\nu p K_1+MpK_2),
\end{align*}
where the first two lines are using the inequalities derived above for $v_j$ and $m_j$, next using the fact that the square root is a sub-additive function (i.e., $\sqrt{a+b}<\sqrt{a}+\sqrt{b}$). The last line follows by using the definition of $K_1$ and $K_2$ we previously established.

For a large enough $J$, applying central limit theorem, $\sqrt{J}(\Bar{v} - \mu_v)\sim \mathcal{N}(0,\Sigma_v)$. The rest of the proof follows by applying the same technique as the bound for the variance of $s_j/m_j$, replacing all the $s$ sub-indices with $v$ sub-indices. Finally,
\begin{align*}
    \Var(\lVert \mathbb{E}[\beta\beta^T\mid y]_{\mathcal{T}}\rVert_2 \mid y) &< J^{-1}(2\pi)^{-p}\det(\Sigma^{-1})m_\nu(y)^{-2} (C + 2M_2\lVert\tilde{\beta}^{(2)}\rVert_2 
    +\lVert\tilde{\beta}^{(2)}\rVert^2_2),
\end{align*}
where $C=p K_2 + 2Mp K_3 + M\frac{p(p-1)}{2}K_2K_1
+M^2K_4+M^2p(p-1)K_2^2$, and $M_2 = \nu p K_1+MpK_2$.

\subsection{Proof of Corollary \ref{cor:svd_decomp}} \label{pf:svd_decomp}

Follows from applying Theorem \ref{th:n_means_moments} with variance $\sigma^2d_i^{-2}$ to each of the $r$ components.

\subsection{Proof of Theorem \ref{th:sure_orthog}} \label{pf:sure_orthog}
    
We know by definition that $\mathrm{SURE} = \lVert y-\tilde y \rVert^2_2 +2\sigma^2\sum_{i=1}^n \frac{\partial \tilde y_i}{\partial y_i}$. For the first term, the bias, we can use directly the prediction $\tilde y = Z \tilde \gamma = Z \mathbb{E}[\gamma\mid \hat \gamma,\sigma^2D^{-2}]$. For the ``degrees of freedom'', denote with $\eta = \frac{d}{d\hat\gamma}\frac{d}{d\hat\gamma^T}\log(m_\nu(\hat\gamma))$. By Proposition 1 of \citet{GriffinBrown}, which is a consequence of Tweedie's formula \citep[][Equation 1.4]{Efron2011tweedie}: 
\begin{align*}
    \Var(\gamma\mid \hat \gamma) = & \sigma^2 (Z^TZ)^{-1} + \sigma^4 (Z^TZ)^{-1}\eta (Z^TZ)^{-1}\\
    = & \sigma^2 D^{-2}(I + \sigma^2 \eta D^{-2}),\\
    \mathrm{tr}(\Var(\gamma\mid \hat \gamma)) = & \sum_i \sigma^2 d_i^{-2}(1+ \sigma^2d_i^{-2}\frac{\partial^2}{\partial\hat\gamma_i^2}\log(m(\hat\gamma)))\\
    = & \sum_i \Var(\gamma\mid \hat \gamma)_{ii}.
\end{align*}
Since we only look into the diagonal terms, we can use the trace to summarize it nicely. We have:
\begin{align*}
    2\sigma^2\sum_{i=1}^n \frac{d \tilde y_i}{d y_i} = & 2\sigma^2 \mathrm{tr}\Big(\frac{\partial \tilde y}{\partial y}\Big)\\
    = & 2\sigma^2\mathrm{tr}\left(Z D^{-1}U^T + \frac{\partial\,\sigma^2 UD^{-1} \nabla_{\hat\gamma}\log(m(\hat\gamma))}{\partial \hat\gamma} \frac{\partial \hat\gamma}{\partial y}\right)\\
    = & 2\sigma^2r +2\sigma^4 \mathrm{tr}(D^{-2}\eta)\\
    = & 2\sigma^2r +2\sigma^4 \sum_{i=1}^r d_i^{-2}\frac{\partial^2}{\partial\hat\gamma^2_i}\log(m(\hat\gamma))\\
    = & 2\sigma^2\sum_{i=1}^r \Var(\gamma\mid \hat\gamma)_{ii}d^2_{i},
\end{align*}
which concludes the proof.

\subsection{Proof of Theorem~\ref{th:sure}:} \label{pf:sure_multivariate}
We know by definition that: $\mathrm{SURE}={\lVert y -\tilde y \rVert_2^2} + {2 \sigma^2\sum_{i=1}^n (\partial \tilde y_i)/(\partial y_i)}$. As in Appendix~\ref{pf:sure_orthog}, we use the prediction of $\tilde y = X\tilde \beta$. Using $\sum_{i=1}^n(\partial \tilde y_i)/(\partial y_i) = \mathrm{tr}((\partial \tilde y)/(\partial y^T))$, it suffices to find an expression for $(\partial \tilde y)/(\partial y^T)$. By definition of $\tilde y$ we have that:
\begin{align}
    \frac{\partial \tilde y}{\partial y^T} = & X \frac{\partial \tilde \beta}{\partial y^T},\nonumber\\
    = & X\frac{1}{m(y)} \frac{\partial  }{\partial y^T}\int_{\mathbb{R}}\beta p(y\mid X,\beta)\pi(\beta)d\beta -\tilde \beta \frac{1}{m(y)}\frac{\partial }{\partial y^T}m(y)\nonumber\\
    = & X \frac{1}{m(y)}\int_{\mathbb{R}^p}-\beta (y-X\beta)^T p(y\mid X, \beta) \pi(\beta)d\beta\Sigma^{-1}\nonumber \\
    & - X\tilde\beta \frac{1}{m(y)}\int_{\mathbb{R}} -(y-X\beta)^Tp(y\mid X,\beta)\pi(\beta)d\beta\Sigma^{-1}  \nonumber\\
    = & X\mathbb{E}[\beta\beta^T]X^T\Sigma^{-1}- X\tilde \beta y^T  + X\tilde\beta y^T-X\tilde\beta\tilde\beta^TX^T\Sigma^{-1} \nonumber\\
    = & X\mathbb{E}[\beta\beta^T\mid y]X^T\Sigma^{-1} - X\tilde\beta\tilde\beta^TX^T\Sigma^{-1},\label{eq:sure_last_line1}
\end{align} 
which is a multivariate version of Tweedie's formula \citep[][Equation 1.4]{Efron2011tweedie}.
Using the variance formulas: $\Var(X\beta\mid y) = X\Var(\beta\mid y)X^T= X\mathbb{E}[\beta\beta^{T}\mid y]X^T-X\tilde\beta\tilde\beta^TX^T$, Equation \eqref{eq:sure_last_line1} becomes:
\begin{align*}
    \frac{\partial \tilde y }{\partial y^T} = & \Var(X\beta\mid y)\Sigma^{-1},
\end{align*}
as required.

\subsection{Proof of Corollary \ref{cor:hats_are_right}}\label{pf:hats_are_right}

Let $p_j = p_\nu(y\mid \mathbf{T}_j)$, and denote with ${p_* = \max_j p_j}$. We want to prove that ${w_j=p_j/(\sum_j p_j)}$. By definition: 
\begin{align*}
  w_j &= \frac{w_j^*}{\sum_{j}w_j^*}\\
  &= \frac{p_j p_{*}^{-1}}{\sum_j p_jp_{*}^{-1}}\\
  &= \frac{p_j}{\sum_j p_j},
\end{align*}
where the second line follows by definition of $w_j^*$, and the third line by cancelling $p_*^{-1}$ in the numerator and denominator. This means that $m_\nu(y)_\mathcal{T}=\sum_{j}p_j$. The first equality we wanted to prove follows by multiplying $\mathbb{E}[\beta\mid y]_\mathcal{T}$ by $X$.

Now, the second equality follows by taking the $X$ out of the variance, and applying the variance formula:

\begin{align*}
\widehat{\Var}(X\beta\mid y) & = X\widehat{\Var}(X\beta\mid y)X^T \\
&=X(\mathbb{E}[\beta\beta^T\mid y]_\mathcal{T} - \mathbb{E}[\beta\mid y]_\mathcal{T}\mathbb{E}[\beta\mid y]_\mathcal{T}^T)X^T,
\end{align*}
concluding the proof.

\bibliographystyle{apalike}
\bibliography{sample}
\onecolumn
\pagestyle{plain}
\setcounter{page}{0}
\renewcommand\thepage{S.\arabic{page}} 
\setcounter{page}{1}

\setcounter{table}{0}
\renewcommand{\thetable}{S\arabic{table}}%
\setcounter{figure}{0}
\renewcommand{\thefigure}{S\arabic{figure}}
\setcounter{section}{0}
\renewcommand{\thesection}{S.\arabic{section}}

\begin{center}
{\Large \textbf{Supplementary Material for} \\\emph{SURE-tuned Bridge Regression} \\\textbf{by}\\ \emph{Jorge Lor\'ia and Anindya Bhadra}}\\
\end{center}
\vspace{1cm}

\section{Simulations with Varying Correlations among the Columns of the Design Matrix}\label{subsec:low_cors}
\begin{figure}[!htb]
    \centering
    \includegraphics[width=15cm,height=7cm]{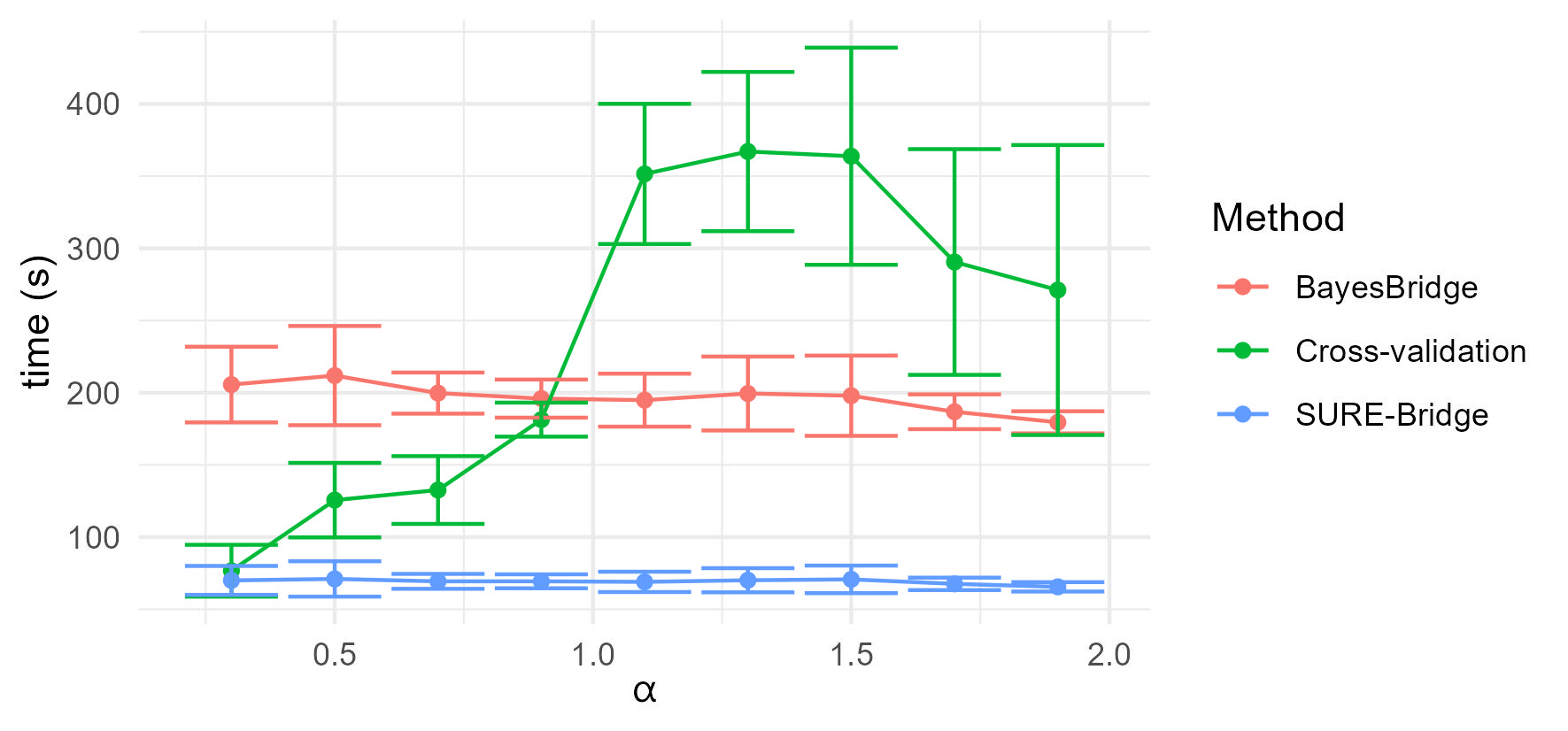}
    \caption{Comparison of average running time (s) $\pm$ SD by method, when changing the $\alpha$ parameter. Using $p=1000$, $n=100$, in design matrices generated using $\rho = 0.1$}
    \label{fig:rho01}
\end{figure}

\begin{figure}[!htb]
    \centering
    \includegraphics[width=15cm,height=7cm]{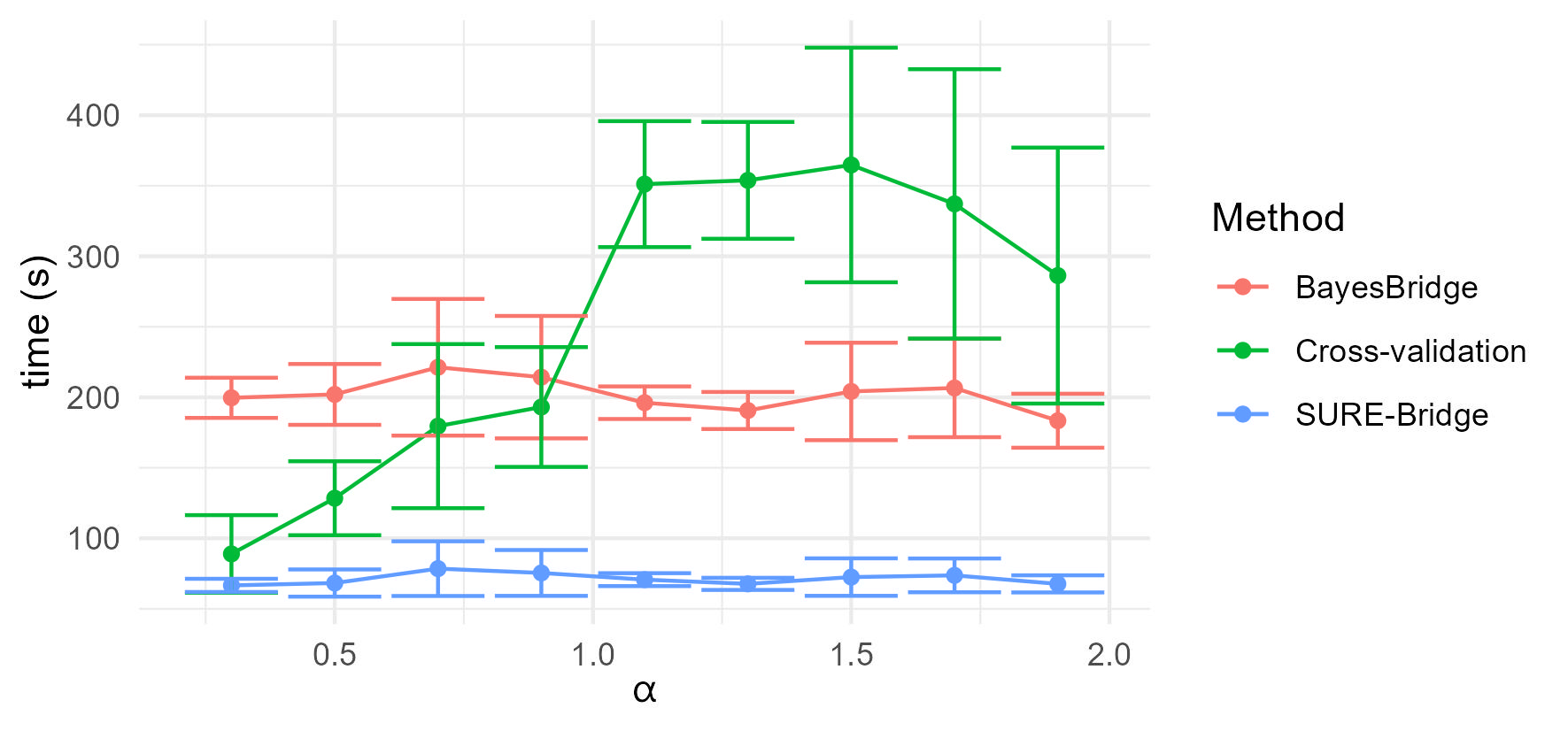}
    \caption{Comparison of average running time (s) $\pm$ SD by method, when changing the $\alpha$ parameter. Using $p=1000$, $n=100$, in design matrices generated using $\rho = 0.5$.}
    \label{fig:rho05}
\end{figure}

We provide additional simulation results with $n=100,\; p=1000$ for $\rho=0.1,0.5$, where $\rho$ is the correlation among the columns of the design matrix $X$. Figures~\ref{fig:rho01} and~\ref{fig:rho05} make the computational advantages of  SURE-Bridge over the competing methods explicit in terms of running time. Next, Tables~\ref{tab:rho01} and~\ref{tab:rho05} display that this computational advantage does not come by sacrificing statistical performance, and the cross validation method performs poorly in regards to SSE.

\begin{table*}[!htb]
\centering
\caption{Average SSE (SD) by method in one hundred out of sample simulated datasets, by $\alpha$. Using $p=1000$, $n=100$, in a design matrix generated with $\rho=0.1$.}\label{tab:rho01}
\begin{tabular}{rllll}
  \hline
$\alpha$ & SURE & SURE-Bridge & BayesBridge & cross validation \\ 
  \hline
  0.30 & 199.87 (0.04) & 202.44 (28.01) & 195.21 (26.21) & 447.24 (145.24) \\ 
  0.50 & 199.86 (0.03) & 206.43 (28.80) & 204.66 (28.45) & 773.52 (551.38) \\ 
  0.70 & 199.86 (0.03) & 195.58 (25.57) & 195.54 (25.57) & 755.39 (265.3) \\ 
  0.90 & 199.86 (0.02) & 199.97 (34.47) & 199.91 (34.47) & 800.74 (225.33) \\ 
  1.10 & 199.87 (0.02) & 198.22 (28.72) & 198.25 (28.75) & 201.28 (42.41) \\ 
  1.30 & 199.87 (0.02) & 199.53 (28.79) & 199.57 (28.78) & 201.49 (36.10) \\ 
  1.50 & 199.87 (0.02) & 196.22 (29.52) & 196.23 (29.54) & 627.28 (2461.64) \\ 
  1.70 & 199.88 (0.02) & 196.85 (28.21) & 196.85 (28.24) & 3095.95 (11037.11) \\ 
  1.90 & 199.88 (0.02) & 201.88 (33.12) & 201.86 (33.11) & 8131.94 (19833.97) \\ 
   \hline
\end{tabular}
\end{table*}

\begin{table*}[!htb]
\centering
\caption{Average SSE (SD) by method in one hundred out of sample simulated datasets, by $\alpha$. Using $p=1000$, $n=100$, in a design matrix generated with $\rho=0.5$}
\label{tab:rho05}
\begin{tabular}{rllll}
  \hline
$\alpha$ & SURE & SURE-Bridge & BayesBridge & cross validation \\ 
  \hline
  0.30 & 199.77 (0.05) & 205.29 (29.87) & 193.84 (27.90) & 344.67 (112.95) \\ 
  0.50 & 199.75 (0.06) & 198.47 (29.23) & 195.69 (28.75) & 515.83 (503.85) \\ 
  0.70 & 199.74 (0.05) & 198.38 (25.71) & 198.32 (25.78) & 477.83 (153.63) \\ 
  0.90 & 199.75 (0.04) & 203.47 (27.90) & 203.46 (27.89) & 524.75 (128.41) \\ 
  1.10 & 199.76 (0.03) & 200.25 (27.80) & 200.29 (27.83) & 200.33 (27.73) \\ 
  1.30 & 199.77 (0.03) & 200.09 (26.24) & 200.14 (26.30) & 213.35 (117.09) \\ 
  1.50 & 199.77 (0.03) & 197.99 (26.16) & 198.00 (26.18) & 602.09 (2892.13) \\ 
  1.70 & 199.78 (0.03) & 199.75 (29.12) & 199.76 (29.16) & 1904.76 (5699.6) \\ 
  1.90 & 199.78 (0.03) & 204.42 (29.29) & 204.46 (29.25) & 3720.81 (8400.79) \\ 
   \hline
\end{tabular}
\end{table*}

\clearpage
\section{Scaling of Computational Times with \texorpdfstring{$n$}{n} %
and \texorpdfstring{$p$}{p}%
}\label{subsec:var_n_p} 
\begin{figure}[!htb]
    \centering
    \includegraphics[width=15cm,height=7cm]{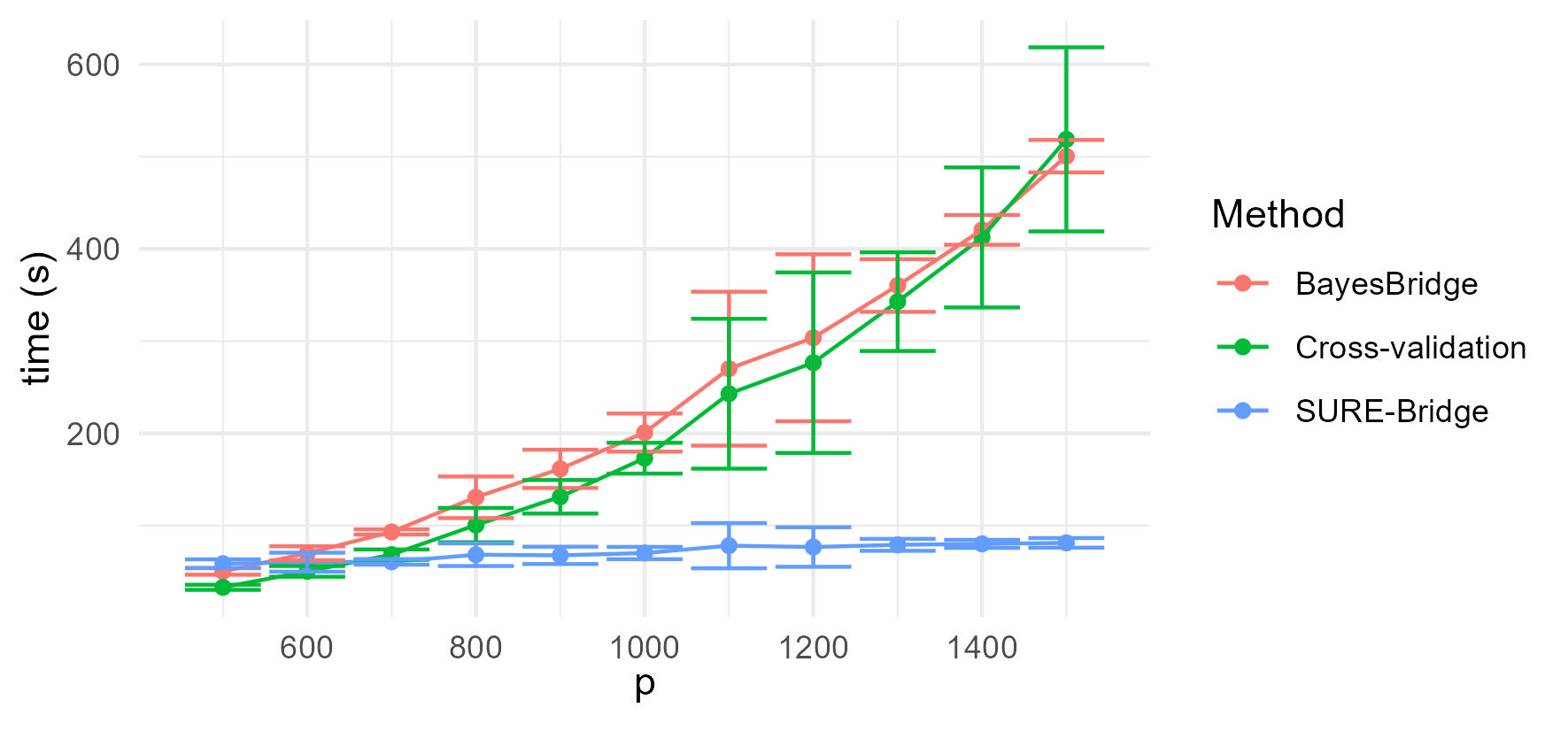}
    \caption{Comparison of average running time (s) $\pm$ SD by method, for a fixed $\alpha=0.7$, varying the number of covariates ($p$), with a fixed number of observations $n=100$ in design matrices generated using $\rho = 0.9$, using a number of signals equal to $\max(0.01p,10)$.}
    \label{fig:var_ps}
\end{figure}

\begin{figure}[!htb]
    \centering
    \includegraphics[width=15cm,height=7cm]{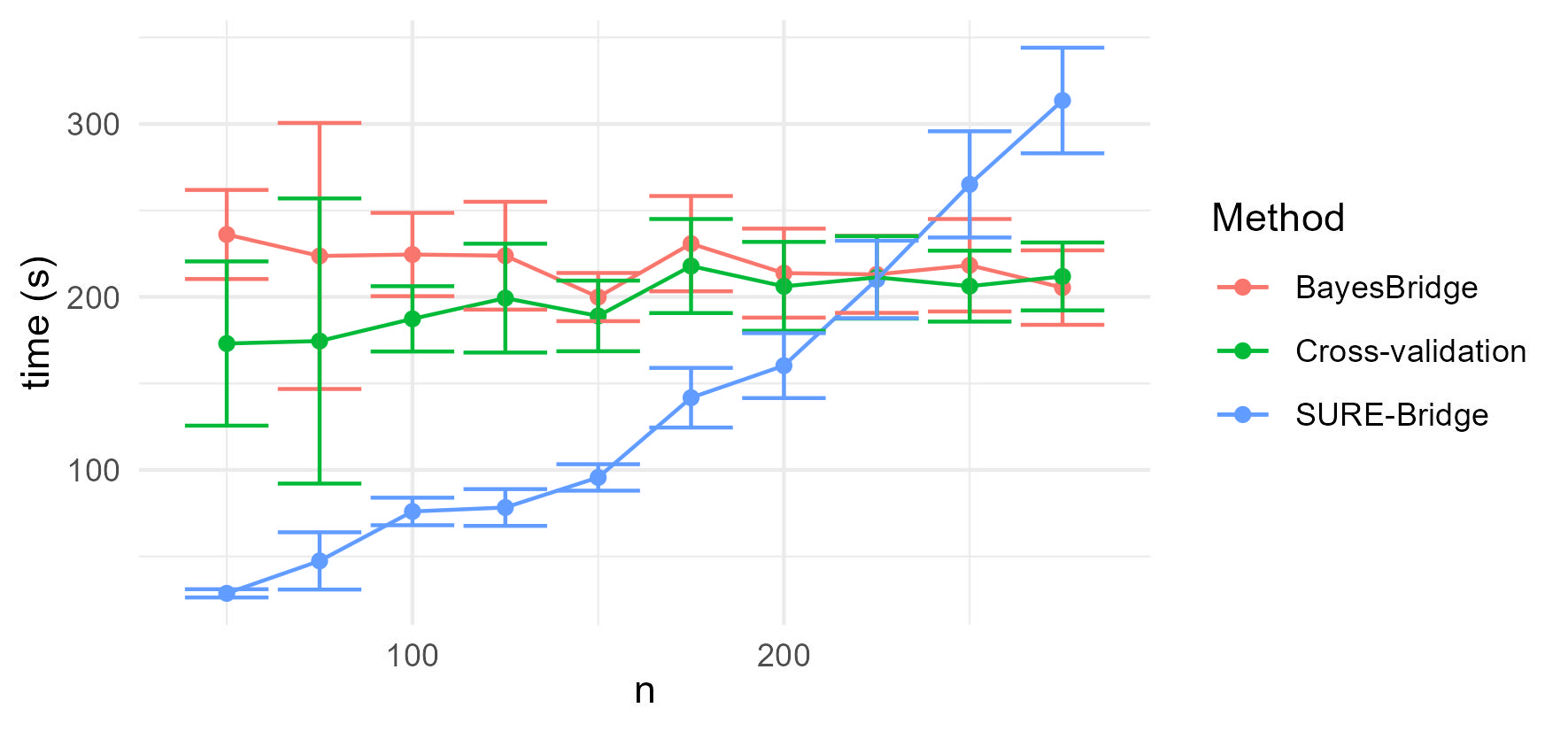}
    \caption{Comparison of average running time (s) $\pm$ SD by method, for a fixed $\alpha=0.7$, varying the number of observations ($n$), with a fixed number of covariates $p=1000$ in design matrices generated using $\rho = 0.9$.}
    \label{fig:varying_ns}
\end{figure}

We present results on the scaling of computational times for all competing approaches.  Figure~\ref{fig:var_ps} displays the scaling of computation times over the number of covariates $p$, for a fixed sample size $n=100$ and Table~\ref{tab:var_ps} displays the corresponding SSEs for the three methods. Similarly, Figure~\ref{fig:varying_ns} displays the scaling of computation times over $n$, for a fixed $p=1000$ and Table~\ref{tab:varying_ns} displays the corresponding SSEs. The overall finding is that the proposed approach (SURE-Bridge) enjoys very favorable scaling for increasing $p$ compared to other methods, explained by its computational complexity as reported in Section~\ref{subsec:compcomplexity}, which is linear in $p$ for a given $n$.
\begin{table*}[!htb]
\centering
\caption{Average SSE (SD) by method in one hundred out of sample simulated datasets, by $p$. Using $\alpha=0.7$, $n=100$, with design matrices generated with $\rho=0.9$, using a number of signals equal to $\max(0.01p,10)$.}
\label{tab:var_ps}
\begin{tabular}{rllll}
  \hline
$p$ & $\mathrm{SURE}$ & SURE-Bridge & BayesBridge & cross validation \\ 
  \hline
   500 & 198.49 (0.34) & 194.35 (25.99) & 190.34 (25.42) & 162.05 (27.71) \\ 
   600 & 198.54 (0.30) & 197.41 (26.59) & 195.44 (26.35) & 166.16 (25.37) \\ 
   700 & 198.62 (0.25) & 198.39 (28.46) & 197.31 (28.36) & 182.47 (25.92) \\ 
   800 & 198.65 (0.28) & 209.46 (32.78) & 209.00 (32.58) & 190.44 (25.03) \\ 
   900 & 198.73 (0.23) & 196.29 (27.53) & 195.87 (27.67) & 189.56 (27.96) \\ 
  1000 & 198.78 (0.23) & 196.30 (30.80) & 196.08 (30.58) & 200.92 (31.23) \\ 
  1100 & 198.92 (0.21) & 200.12 (30.22) & 199.93 (30.30) & 207.55 (37.69) \\ 
  1200 & 199.03 (0.18) & 198.33 (26.94) & 198.26 (26.99) & 209.31 (62.94) \\ 
  1300 & 199.12 (0.15) & 194.35 (31.62) & 194.22 (31.51) & 204.18 (43.43) \\ 
   \hline
\end{tabular}
\end{table*}

\begin{table*}[!htb]
\centering
\caption{Average SSE (SD) by method in one hundred out of sample simulated datasets, by $n$. Using $\alpha=0.7$, $p=1000$, with design matrices generated with $\rho=0.9$.}
\label{tab:varying_ns}
\begin{tabular}{rllll}
  \hline
$n$ & $\mathrm{SURE}$ & SURE-Bridge & BayesBridge & cross validation \\ 
  \hline
   50.00 &  99.46 (0.13) &  97.03 (17.80) &  97.03 (17.85) & 205.37 (340.94) \\ 
   75.00 & 149.16 (0.15) & 145.72 (21.95) & 145.70 (21.92) & 147.09 (46.46) \\ 
  100.00 & 198.78 (0.23) & 196.30 (30.80) & 196.08 (30.58) & 200.92 (31.23) \\ 
  125.00 & 248.37 (0.32) & 248.52 (28.22) & 245.70 (28.16) & 247.08 (35.53) \\ 
  150.00 & 297.93 (0.36) & 305.16 (37.37) & 294.23 (36.24) & 295.97 (32.54) \\ 
  175.00 & 347.44 (0.47) & 349.75 (42.60) & 329.95 (40.54) & 346.02 (42.03) \\ 
  200.00 & 396.92 (0.51) & 390.32 (36.19) & 361.42 (33.87) & 381.55 (45.25) \\ 
  225.00 & 446.31 (0.67) & 450.82 (40.57) & 409.33 (36.50) & 430.00 (46.36) \\ 
  250.00 & 495.75 (0.85) & 493.38 (41.90) & 439.73 (37.78) & 476.27 (45.74) \\ 
  275.00 & 544.96 (0.84) & 552.56 (43.92) & 486.33 (39.81) & 529.72 (50.35) \\ 
   \hline
\end{tabular}
\end{table*}

\clearpage
\section{Simulation Results under Deviations from Model Assumptions} \label{subsec:devs} 
\begin{figure*}[!htb]
   \centering
   \includegraphics[height=7cm,width=15cm]{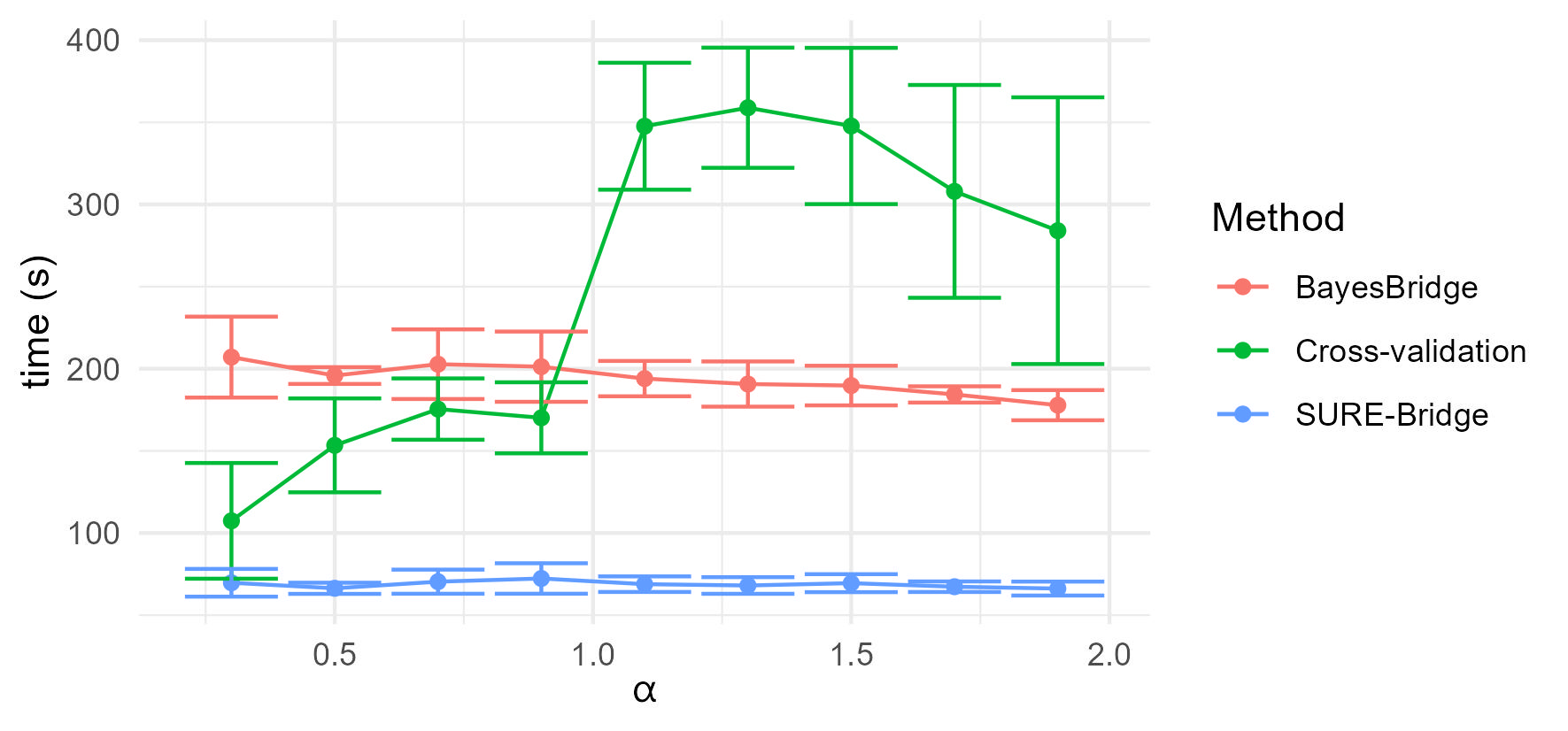}
   \caption{Comparison of average running time (s) $\pm$ SD by method, when changing the $\alpha$ parameter. Using  $n=100$, $p=1000$, in matrices generated using $\rho = 0.9$, with $t$-distributed errors with three degrees of freedom and variance one.}
   \label{fig:line_plot_dev_heavy_n_100_p_1000}
\end{figure*}

\begin{figure*}[!htb]
   \centering
   \includegraphics[height=7cm,width=15cm]{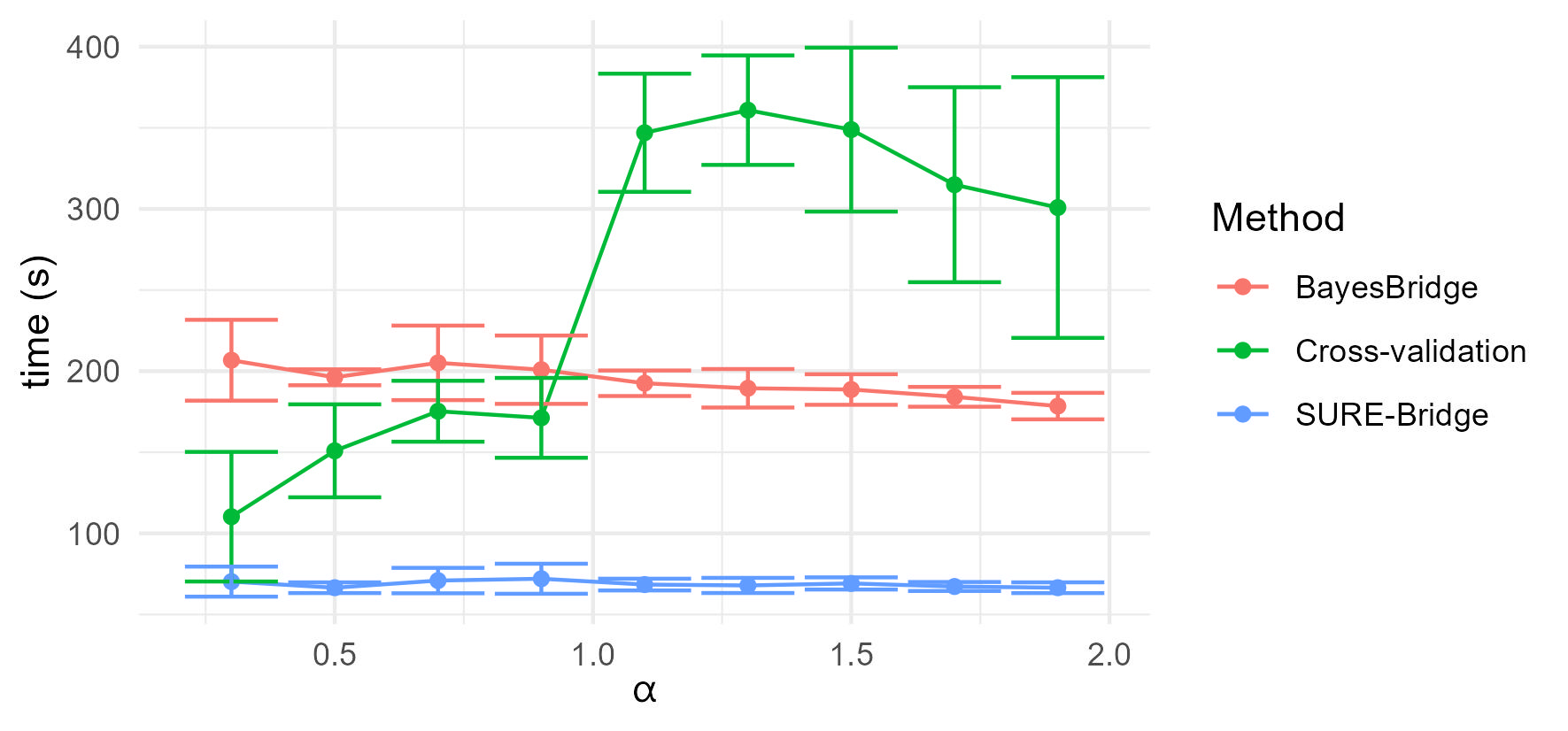}
   \caption{Comparison of average running time (s) $\pm$ SD by method, when changing the $\alpha$ parameter. Using  $n=100$, $p=1000$, in matrices generated using $\rho = 0.9$, with correlated error terms.}
   \label{fig:line_plot_dev_corr_errs_n_100_p_1000}
\end{figure*}

We perform robustness checks under two different deviations from our modeling assumptions: normality and uncorrelatedness of the error terms.  In Figure~\ref{fig:line_plot_dev_heavy_n_100_p_1000} and Table~\ref{tab:t_dist_errors}, we report results on noise generated from a $t$-distribution with three degrees of freedom and variance one, violating the assumption of normality. Similarly, Figure~\ref{fig:line_plot_dev_corr_errs_n_100_p_1000} and Table~\ref{tab:cor_errors} display results on noise terms generated from a multivariate Gaussian with variance one and correlations equal to $0.1$, violating the assumption of uncorrelated errors. We remark that Stein's unbiased risk estimate is no longer an unbiased estimate of the out of sample SSE under these deviations. Remarkably, SURE is still within one standard deviation of the obtained SSEs of the SURE-Bridge method. However, the standard deviations of the SSEs are larger than those displayed in Section~\ref{sec:Results}.

\begin{table*}[!htb]
\centering
\caption{Average SSE (SD) by method in one hundred out of sample simulated datasets, by $\alpha$. Using $p=1000$, $n=100$, in a matrix generated with $\rho=0.9$, with t-distributed errors with three degrees of freedom and variance one.}
\label{tab:t_dist_errors}
\begin{tabular}{rllll}
  \hline
$\alpha$ & SURE & SURE-Bridge & BayesBridge & cross validation \\ 
  \hline
  0.30 & 198.82 (0.35) & 189.88 (87.38)  & 162.18 (78.31)  & 187.50 (66.13) \\ 
  0.50 & 198.77 (0.28) & 205.37 (87.28)  & 197.12 (85.69)  & 195.93 (66.35) \\ 
  0.70 & 198.75 (0.26) & 199.80 (84.73)  & 199.61 (84.56)  & 195.79 (57.52) \\ 
  0.90 & 198.79 (0.20) & 207.86 (116.03) & 207.77 (116.09) & 217.25 (93.39) \\ 
  1.10 & 198.83 (0.17) & 208.50 (85.00)  & 208.41 (85.04)  & 206.96 (77.41) \\ 
  1.30 & 198.87 (0.17) & 208.34 (100.16) & 208.33 (100.24) & 213.31 (113.72) \\ 
  1.50 & 198.89 (0.17) & 198.38 (88.02)  & 198.41 (88.00)  & 278.59 (678.51) \\ 
  1.70 & 198.90 (0.17) & 206.04 (90.98)  & 206.04 (90.96)  & 510.25 (1078.58) \\ 
  1.90 & 198.90 (0.17) & 212.02 (110.17) & 211.98 (110.14) & 1133.25 (2016.05) \\ 
   \hline
\end{tabular}
\end{table*}

\begin{table*}[!htb]
\centering
\caption{Average SSE (SD) by method in one hundred out of sample simulated datasets, by $\alpha$. Using $p=1000$, $n=100$, in a matrix generated with $\rho=0.9$, using Gaussian errors generated with all their correlations equal to $0.1$.}
\label{tab:cor_errors}
\begin{tabular}{rllll}
  \hline
$\alpha$ & SURE & SURE-Bridge & BayesBridge & cross validation \\ 
  \hline
0.30 & 198.88 (0.3) & 196.57 (35.57) & 168.96 (31.96) & 198.19 (49.44) \\ 
  0.50 & 198.78 (0.27) & 202.37 (39.25) & 194.27 (38.21) & 199.68 (36.7) \\ 
  0.70 & 198.76 (0.24) & 201.6 (35.2) & 201.21 (35.12) & 204.01 (34.55) \\ 
  0.90 & 198.79 (0.19) & 194.73 (29.51) & 194.67 (29.34) & 216.93 (44.75) \\ 
  1.10 & 198.83 (0.17) & 206 (44.82) & 205.93 (44.78) & 206.33 (44.39) \\ 
  1.30 & 198.87 (0.17) & 203.01 (44.51) & 203.02 (44.64) & 204.36 (45.04) \\ 
  1.50 & 198.88 (0.17) & 202.32 (42.37) & 202.34 (42.43) & 245.54 (317.54) \\ 
  1.70 & 198.89 (0.17) & 200.15 (41.08) & 200.19 (41.05) & 448.14 (794.65) \\ 
  1.90 & 198.9 (0.17) & 196.38 (40.27) & 196.42 (40.26) & 948.17 (1864.85) \\ 
   \hline
\end{tabular}
\end{table*}

\section{Numerical Comparison of the Efficiency of the Estimates} \label{subsec:efficiency}
\begin{figure}[!htb]
    \centering
       \includegraphics[height=7cm,width=15cm]{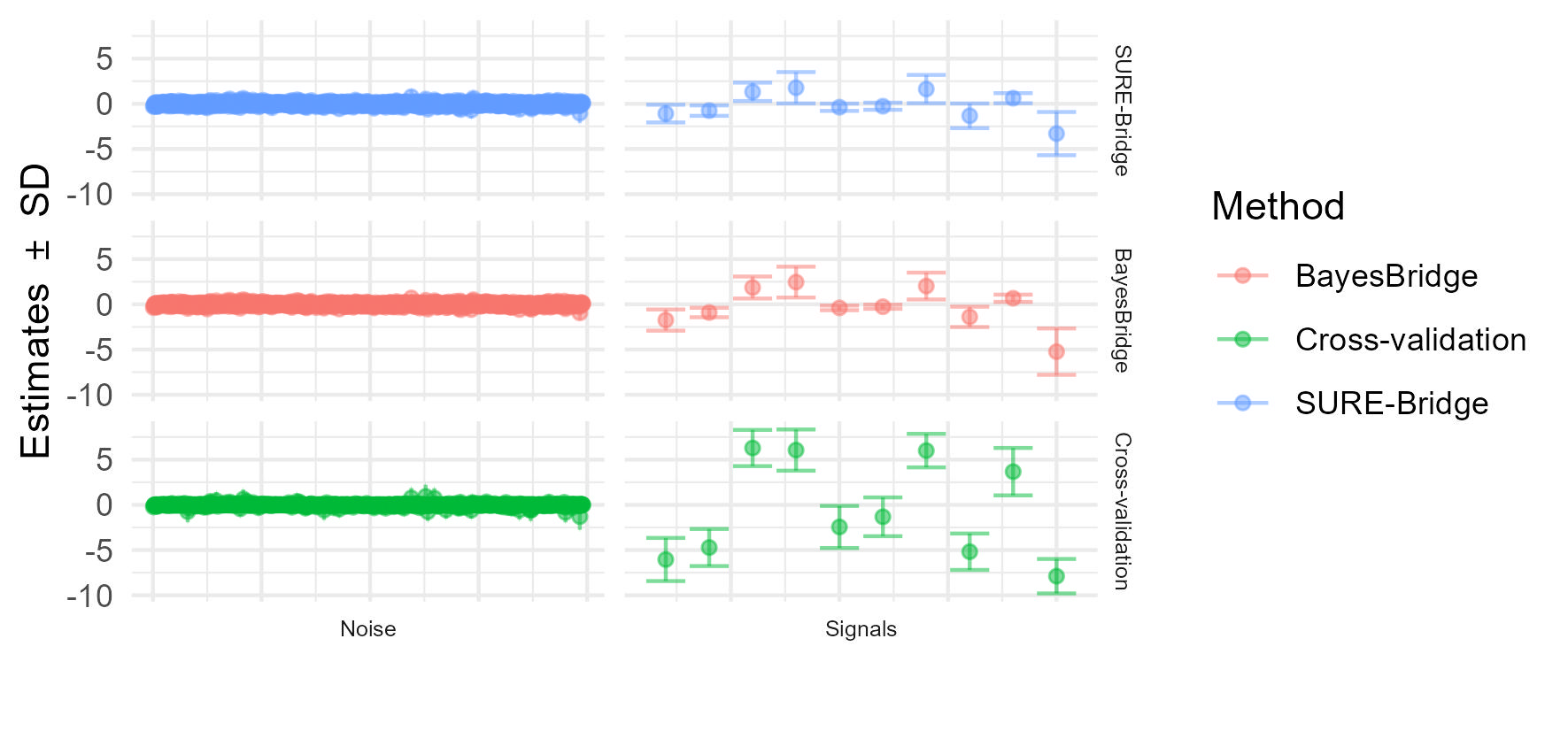}
    \caption{Estimates $\pm$ SD from one hundred bootstrap replicates, by method, using $\alpha=0.5$ for ${n=100},{p=1000}$ and $\rho=0.9$. The signals correspond to the last ten entries.}
    \label{fig:boot}
\end{figure}

An exact formula for the standard error of our estimates is unavailable. Hence, we use bootstrap to assess the efficiency of the estimates, with the caveat that a theoretical investigation of the consistency of bootstrap for this model is beyond the scope of the current paper. Table~\ref{tab:boot} shows the average squared deviation of the estimates by method using $B=100$ bootstrap replicates. Figure~\ref{fig:boot} shows the bootstrap estimates of the three methods for $\alpha=0.5$. We find that our method performs similarly in signal recovery compared to BayesBridge. Cross validation typically results in less bias at recovering the signals in the training set in our experiments. However, as seen in the main paper, this typically leads to poorer prediction performances, which is not surprising in the light of the role of bias--variance trade-off for out of sample prediction.
\begin{table*}[!htb]
\centering
\caption{Bootstrap mean squared error  estimates by $\alpha$ and method, using $p=1000$, $n=100$, in a design matrix generated with $\rho=0.9$.}
\label{tab:boot}
\begin{tabular}{rrrr}
  \hline
$\alpha$ & SURE-Bridge & BayesBridge & cross validation \\ 
  \hline
0.30 & 0.97 & 0.53 & 0.45 \\ 
  0.50 & 0.89 & 0.78 & 0.44 \\ 
  0.70 & 0.89 & 0.87 & 0.40 \\ 
  0.90 & 0.91 & 0.91 & 0.35 \\ 
  1.10 & 0.93 & 0.93 & 0.53 \\ 
  1.30 & 0.93 & 0.93 & 0.75 \\ 
  1.50 & 0.94 & 0.94 & 0.86 \\ 
  1.70 & 0.94 & 0.94 & 0.91 \\ 
  1.90 & 0.94 & 0.94 & 0.94 \\ 
   \hline
\end{tabular}
\end{table*}

\section{Non-uniform Running Times for Cross Validation over \texorpdfstring{$\alpha$}{alpha}%
}\label{subsec:cross validation}
\begin{figure}[!htb]
    \centering
    \includegraphics[height=5.5cm,width=15cm]{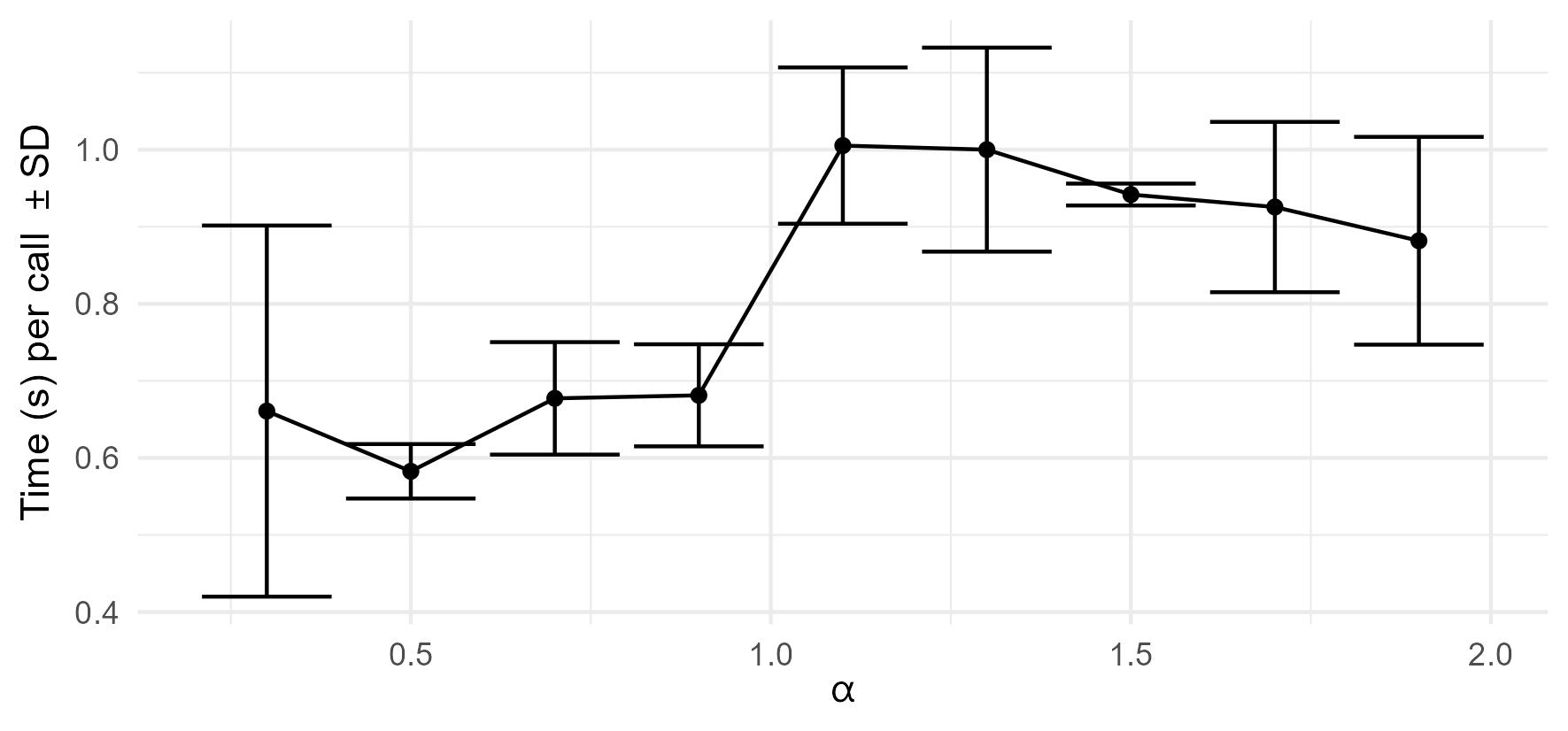}
    \caption{Average time per call ($\pm SD$) of the EM procedure for each $\alpha$ in the simulation of Section~\ref{sec:Results}.}
    \label{fig:nonuniform_CV}
\end{figure}

\begin{figure}[!htb]
    \centering
    \includegraphics[height=5.5cm,width=15cm]{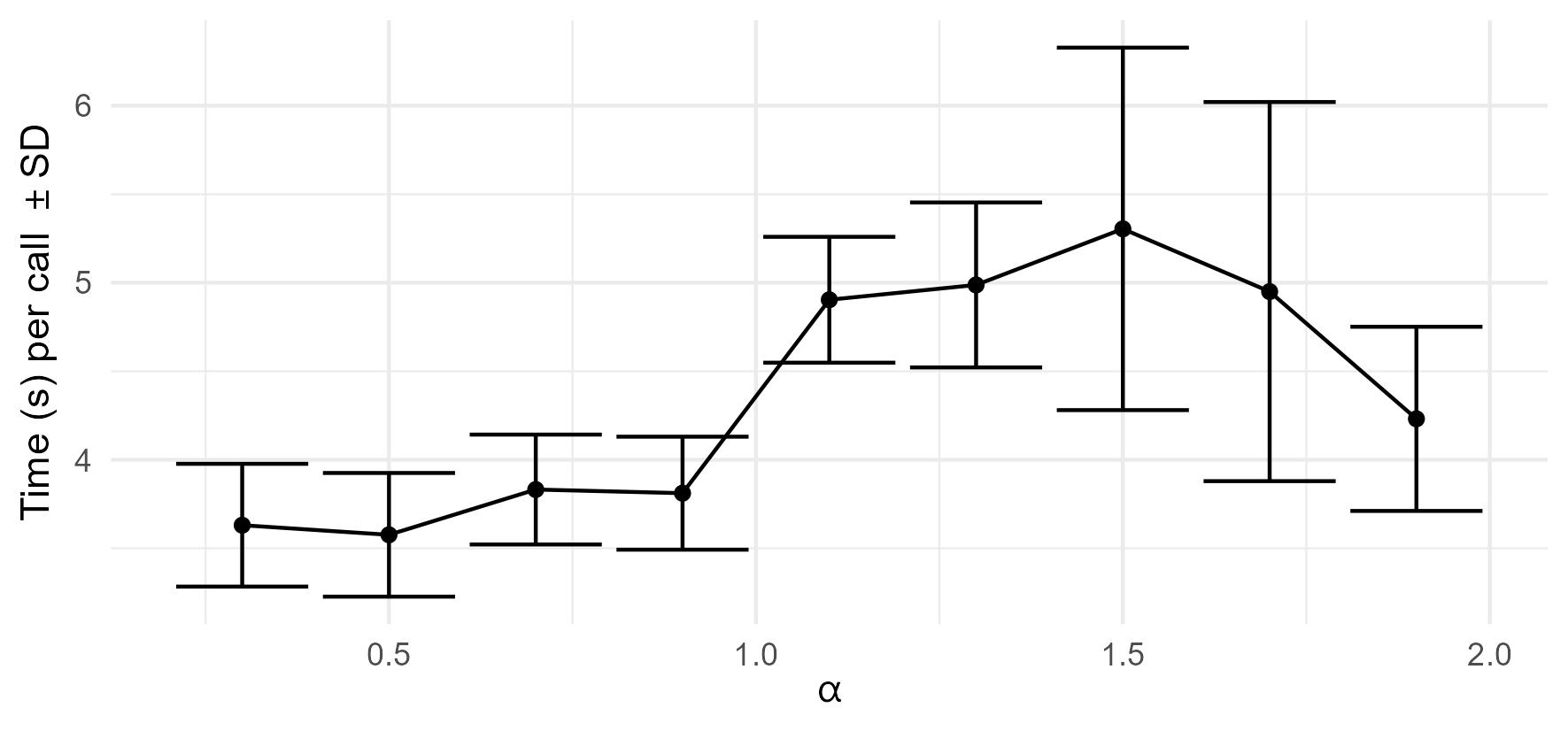}
    \caption{Average time per call ($\pm SD$) of the EM procedure for each $\alpha$ in Section~\ref{sec:real_data}.}
    \label{fig:nonuniform_CV_photos}
\end{figure}

The figures in the main paper as well as the rest of the Supplementary Material have a non-uniform running time across $\alpha$ for the cross validation method which uses the EM procedure by \citet{PolsonScottEM}, with values of $\alpha\le1$ typically resulting in faster computation. Figures~\ref{fig:nonuniform_CV} and~\ref{fig:nonuniform_CV_photos} display the average time per call of the EM procedure against $\alpha$. 
The reason is when the regression coefficient is very close to zero, \citet{PolsonScottEM} drop that coefficient, and remove the corresponding column of the design matrix, which results in computational savings. 
In the context of Bridge regression~(Eq.~\eqref{eq:bridge}), \citet{FanLiSCADJASA} prove that when $\alpha > 1$ there is no guarantee of sparsity, in contrast to $\alpha\leq 1$, which does give sparse estimates. The implementation of the algorithm by \citet{PolsonScottEM} indeed takes advantage of these zero-ed entries by removing them from the problem, typically resulting in faster solutions for $\alpha<1$. 

\section{Numerical Verification of Variance Bounds}
To avoid computing the constants that bound the variances in Theorem \ref{th:finite_var}, we compare the ratios of the variances when changing the number of Monte Carlo samples $J$. Specifically, we compute the quantities of interest in $100$ independent simulations, using $n=20,p=50,\rho = 0.3$, with 5 signals of 10 and the rest centered standard normals. We do this with three different number of samples: $J=10,100,1000$. Next we compute the variance of the quantities, and take the ratios. We display the results in Table~\ref{tab:variance_comparison}. The first column denotes the ratio between which variances according to their number of samples, the second column denotes the ratio of $\lVert\mathbb{E}[\beta\mid y]_\mathcal{T}\rVert_2$, and the third one this ratio for $\lVert\Var[\beta\mid y]_\mathcal{T}\rVert_2$. Theorem~\ref{th:finite_var_vector} indicates that we would expect these ratios to be around $1/10$ for one order of magnitude increase in $J$, which the results confirm.
\begin{table}[!htb]
\centering
\begin{tabular}{rrr}
  \hline
Ratio & $\lVert\mathbb{E}[\beta\mid y]_\mathcal{T}\rVert_2$ &  $\lVert\Var[\beta\mid y]_\mathcal{T}\rVert_2$\\ 
  \hline
$v_{10^2}/v_{10}$ & 0.40 & 0.26 \\ 
$v_{10^3}/v_{10^2}$ & 0.08 & 0.12 \\ 
$v_{10^4}/v_{10^3}$ & 0.13 & 0.23 \\ 
   \hline
\end{tabular}
\caption{Ratio of variances of the norms\label{tab:variance_comparison}}
\end{table}

\end{document}